\documentclass[%
twocolumn,
 amsmath,amssymb,
aps,
pra,
]{revtex4-1}
\usepackage{listings}
\usepackage{graphicx}% Include figure files
\usepackage{dcolumn}% Align table columns on decimal point
\usepackage{bm}% bold math
\usepackage{natbib}
\usepackage[colorlinks,citecolor=red,linkcolor=blue,urlcolor=blue]{hyperref}
\usepackage{caption}
\usepackage{hhline}
\usepackage{float}
\usepackage{mhchem}
\usepackage[nodisplayskipstretch]{setspace}
\usepackage{calrsfs}
	\DeclareMathAlphabet{\pazocal}{OMS}{zplm}{m}{n}
\usepackage{physics}
\usepackage{appendix}
\usepackage[export]{adjustbox}
\newcommand{\bk}{\bold{k}}

\newcommand{\txdag}[1]{\hat{t}^{x\dagger}_{\bold{r}_i+\boldsymbol{\delta}_{#1}}}

\newcommand{\tydag}[1]{\hat{t}^{y\dagger}_{\bold{r}_i+\boldsymbol{\delta}_{#1}}}

\newcommand{\txdagn}{\hat{t}^{x\dagger}_{\bold{r}_i}}
\newcommand{\txn}{\hat{t}^{x}_{\bold{r}_i}}
\newcommand{\tydagn}{\hat{t}^{y\dagger}_{\bold{r}_i}}
\newcommand{\tyn}{\hat{t}^{y}_{\bold{r}_i}}
\newcommand{\tzdagn}{\hat{t}^{z\dagger}_{\bold{r}_i}}
\newcommand{\tzn}{\hat{t}^{z}_{\bold{r}_i}}

\begin{document}

%\preprint{APS/123-QED}

\title{$U(1)$-Symmetry protected Dirac nodal loops of  triplons in \ce{SrCu2(BO3)2}}% Force line breaks with \\

\author{Dhiman Bhowmick\href{https://orcid.org/0000-0001-7057-1608}{\includegraphics[scale=0.12]{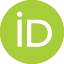}}}
\author{Pinaki Sengupta\href{https://orcid.org/0000-0003-3312-2760}{\includegraphics[scale=0.12]{orcid.png}}}%
\affiliation{%
 School of Physical and Mathematical Sciences, Nanyang Technological University, 21 Nanyang Link, Singapore 637371, Singapore \\
}%

\date{\today}% It is always \today, today,

\begin{abstract}
We demonstrate the appearance of symmetry protected triplon Dirac modal lines in the low energy excitation spectrum of a realistic microscopic model of the geometrically frustrated quantum magnet \ce{SrCu2(BO3)2} in its high symmetry phase. The symmetry-allowed Dzyaloshinskii-Moriya interactions induce
dispersive trilpon bands within the bond-operator formalism that cross linearly
over an extended closed path in the Brillouin zone. Our results establish that 
the nodal lines are protected by a $U(1)$-symmetry and robust against
perturbations that preserve this symmetry. In the presence of a longitudinal field, the nodal loop shrinks and vanishes for a sufficiently strong field. 
\end{abstract}

\pacs{Valid PACS appear here}% PACS, the Physics and Astronomy
\maketitle

%\tableofcontents

\section{\label{sec1}Introduction}
The study of topological phases of matter has grown into one of the most 
active frontiers of contemporary condensed matter physics. Recently, the 
experimental observation of Weyl fermions in \ce{TaAs}\cite{WeylSemiMetal}, 
building upon previous theoretical studies\cite{WeylSemiMetal2,WeylSemiMetal3}, 
has generated widespread interest in topological semimetals. In these systems, a
linear crossing of two bands at one or more points in momentum space is 
topologically protected. Following the discovery of \ce{TaAs}, materials with 
Dirac nodal lines or nodal loops were discovered where the linear band crossing 
exists over an extended region in the Brillouin zone (BZ), thus expanding the 
types of semimetals\cite{NL1,NL2,NL3,NL4,NL5,NL6,NL7,NL8,Mirror3,NL10,NL11,NL12}. 
The nodal lines or loops in these systems is protected by the presence of certain 
symmetries like inversion\cite{NL4,Inversion2,SymmetryDemanded}, 
mirror-reflection\cite{NL3,Mirror2,Mirror3,NL10} or 
glide-plane\cite{GlidePlane1,GlidePlane2,GlidePlane3}. 
Since topological phases are often manifestations of the 
geometry of the band structure which do not depend 
on the quantum statistics of the quasi-particles involved, efforts to realize
bosonic analogs of topological phases have gained interest in the recent past. 
Bosonic counterparts of the topological phenomenons have been explored in wide
ranging bosonic systems such as 
photons\cite{TopologicalPhotonics,TopologicalPhotonics2}, 
exciton-polaritons\cite{TopologicalPolariton,TopologicalPolariton2}, 
 magnons\cite{TopologicalMagnon,TopologicalMagnon2,TopologicalMagnon3,TopologicalMagnon4,TopologicalMagnon5,TopologicalMagnon6,TopologicalMagnon7,new1,new2,new3,new4,new5, WeylMagnon1,WeylMagnon2,WeylMagnon3,WeylMagnon4,WeylMagnon5,WeylMagnon6}, magnon-polarons\cite{MagnonPolaron1,MagnonPolaron2,MagnonPolaron3,MagnonPolaron4}.
Quantum magnets, in particular, offer a promising route to search for novel 
topological phases of quantized magnetic excitations such as magnons and triplons
that obey Bose-Einstein statistics.

The band structure geometry that drives topological properties is largely 
determined by the symmetries of the lattice and the Hamiltonian. While 
fermionic phases are governed by lattice space group symmetries, magnetic systems
are governed by a richer magnetic symmetry group, whose elements consist of 
products of a space group symmetry operation and the time reversal operation.
The geometry of the band structure of magnetic quasiparticle excitations are
governed by elements of the magnetic symmetry group. Dirac magnon nodal lines
that are protected by simultaneous inversion (${\cal P}$) and time reversal 
(${\cal T}$) symmetry have been predicted to appear in anisotropic pyrochlore
ferromagnets\cite{PyrochloreFerromagnet}, the spin-web compound \ce{Cu_3TeO_6}\cite{SpinWeb} and layered honeycomb 
antiferromagnets\cite{LayeredAntiFerromagnet}.

Moreover, the nodal line semimetals are mostly studied theoretically and experimentally in the three dimensional systems. However, recently it has been shown the two dimensional counterpart for Dirac nodal lines also exist\cite{2DSemimetal,2DSemimetal2,2DSemimetal3,2DSemimetal4}. Although as a bosonic  counterpart the presence of Dirac nodal line in the magnetic excitations of a quasi-2D ferromagnetic honeycomb lattice has been investigated\cite{Quasi2D}, but as far as our knowledge no studies show the presence of Dirac nodal line magnetic excitation in completely two-dimensional magnetic system without interlayer coupling.

In this work, we investigate the high symmetry crystal phase of 
the geometrically frustrated quantum magnet \ce{SrCu2(BO3)2} and demonstrate 
the appearance of field 
induced Dirac nodal loops in triplon bands for perfectly 2D-system.  
Previous studies for nodal line magnons have been restricted
to excitations above ground states with classical (static) spin ordering
%were done for the long-range ordered classical 
%ferromagnetic or anti-ferromagnetically ordered states or ground states
\cite{PyrochloreFerromagnet,SpinWeb,LayeredAntiFerromagnet}. In contrast to
those studies, we have explored magnetic excitations in the dimerized phase
of the Shastry-Sutherland model (with additional DM (DM) interactions). To the best of our knowledge, ours is the first study to show that Dirac nodal line also can exists in the excitation spectrum of purely quantum-mechanical ground state with only short range order. The high-symmetry phase of \ce{SrCu2(BO3)2} consists of weakly coupled layers in which the \ce{Cu^{2+}} ions are arranged in
an orthogonal dimer configuration constituting the non-symmorphic 
Shastry-Sutherland lattice\cite{Shastry}.  The space group of \ce{SrCu2(BO3)2}
is $I4/mcm$ in the high symmetry phase (at high temperatures) 
and is reduced to $I42m$ at
$T=395$K via a structural transition\cite{structure1,structure2}. 
The symmetry determines the allowed terms in the Hamiltonian, in particular, 
the nature of Dzyaloshinskii-Moriya (DM) interactions\cite{phase4}. 

The dominant (magnetic) interaction in \ce{SrCu2(BO3)2}
is antiferromagnetic Heisenberg superexchange between \ce{Cu^{2+}} ions
across intra- and inter-dimer bonds. In the high symmetry phase,
DM interaction is allowed only on the inter-dimer bonds with the DM vector
normal to the plane, which preserves the U(1)-symmetry of the spin-system\cite{phase4}. The ground state of \ce{SrCu2(BO3)2} is well approximated
by the dimer singlet 
phase where the spins on each dimer form a singlet\cite{phase1,phase2,phase3,phase4}. In the absence of DM 
interactions, the lowest excitation would consist of 3-fold degenerate 
dispersionless triplons on the dimer bonds. In the presence of DM interactions,
the triplon bands acquire finite dispersion and non-zero Berry curvature. The
nature of DM interactions determine the evolution of the triplon bands in an
external magnetic field. This has been extensively investigated in the
low-symmetry phase of \ce{SrCu2(BO3)2} where upon the application of a longitudinal magnetic field
a gap opens up in the spectrum and the bands acquire a non-zero Chern number\cite{triplon,triplon2,triplon3}.
%(the gap eventually closes and reopens with topologically trivial bands at
%a critical value of the applied field). 
In this work we investigate the 
triplon bands in the high symmetry phase of \ce{SrCu2(BO3)2} in an external longitudinal
magnetic field. Our results show that in this phase, the triplon bands exhibit
Dirac nodal loop which is protected by $U(1)$-symmetry. When a longitudinal field is turned on, the nodal loop
shrinks continually with increasing field and eventually disappears at a critical field
with the opening of a gap. We analyze the mechanism of emergence and 
properties of the Dirac nodal loop within the framework of bond operator
formalism.

\section{\label{sec4} Model and method}
%\subsection*{Microscopic model.}
\begin{figure}[H]
\includegraphics[width=0.5\textwidth]{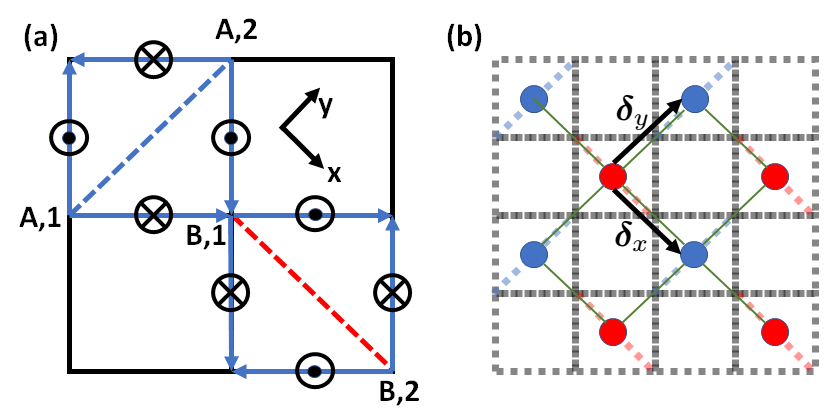} % this command will be ignored
\caption{\textbf{The Shastry-Sutherland lattice.} (a) The schematic of Shastry-Sutherland lattice. The circled-dots(circled-crosses) represent the perpendicular DM-interaction out of(into) the plane of paper. The dotted blue line and dotted red line represent the dimer-bond A and dimer-bond B of the lattice respectively. (b) The effective square lattice structure after bond-operator transformation. The dimer-bonds convert into points and on each point only one quasi-particle triplon can reside. The dimer-A is converted into blue dot and dimer-B is converted into red dot.}
\label{lattice}
\end{figure}

The figure Fig.\ref{lattice}(a) illustrates the Shastry-Sutherland lattice including the DM-interactions present in the high symmetry phase of \ce{SrCu2(BO3)2}. The Hamiltonian of the system is given by,
\begin{align}
\pazocal{H}=&J' \sum_{\left\langle i,j \right\rangle_d} \bold{S}_{i}\cdot\bold{S}_{j} + J \sum_{\left\langle i,j \right\rangle_{x,y}} \bold{S}_{i}\cdot\bold{S}_{j} \nonumber\\
  &+ \bold{D}_\perp\cdot\sum_{\left\langle i,j \right\rangle_{x,y}} \left(\bold{S}_{i}\times\bold{S}_{j}\right)-g_zh_z\sum_{i}\bold{S}_{i}^z,
  \label{eq1}
\end{align}
where $\left\langle ... \right\rangle_{x,y}$, $\left\langle ... \right\rangle_d$ denote the summation over the sites belonging to intra- and inter-dimer bonds respectively. %$i$, $j$ denotes the sites on the Shastry-Sutherland layer. 
$J$ and $J'$ denote the strengths of the corresponding Heisenberg interaction terms. $\bold{D}_\perp$ is the DM-interaction shown as in Fig.\ref{lattice}(a). The fourth term is the Zeeman coupling of the spins with the magnetic field $h_z$ perpendicular to the Shastry-Sutherland layer with Lande-g factor $g_z$. The theoretical studies in Ref.\cite{triplon,triplon2,triplon3} shows a vast phase diagram as well as the topological properties of the magnetic excitations. We show that the presence of $U(1)$-symmetry and the DM-interaction $D_\perp$ in the material give rise to $U(1)$-symmetry protected Dirac-nodal line in the bands of low lying magnetic excitations in the system.

 In the absence of DM-interactions, the possible ground states of the Shastry-Sutherland lattice are dimer phase, N{\' e}el phase and plaquette-singlet phase. The dimer phase of the Shastry-Sutherland lattice exists for $J'\lessapprox 0.7 J$\cite{phase1,phase2,phase3}. The singlet dimer phase of Shastry-Sutherland lattice is experimentally realized in \ce{SrCu(BO3)2}, where the \ce{Cu}$^{2+}$ carries spin-$1/2$ degrees of freedom and forming a orthogonal dimer model\cite{dimer1,structure1,dimer3,dimer4}. The chosen DM-interactions are symmetry allowed for the high symmetry phase of \ce{SrCu2(BO3)2}\cite{structure1,structure2,triplon}. 
 %at a temperature above $395$K We propose that the high symmetry phase of \ce{SrCu2(BO3)2} might be induced by pressure at low temperature. 
 In the presence of DM-interaction the dimer-singlet phase persists as the ground state for a finite  parameter range $-0.2\lessapprox D_\perp/J \lessapprox 0.2$\cite{phase4}. In this study we show that the presence of U(1)-symmetry and perpendicular DM-interaction $D_\perp$ gives rise to Dirac nodal line of the low lying magnetic excitations of this system.  

%\subsection*{Triplon picture.}
The ground state of the canonical Shastry-Sutherland model in the large
$J/J'$ regime (applicable to \ce{SrCu2(BO3)2}) is a product of singlet-dimer states on the red and blue-bonds(Fig.\ref{lattice}(b)). Hence it is natural to use the
bond operator formalism to study the Hamiltonian (\ref{eq1}). The local Hilbert
space on a single dimer consists of a singlet 
$\ket{s}=\left(\ket{\uparrow\downarrow}-\ket{\downarrow\uparrow}\right)/\sqrt{2}$ and
three triplons $\ket{t_x}=i\left(\ket{\uparrow\uparrow}-\ket{\downarrow\downarrow}\right)/\sqrt{2}$, $\ket{t_y}=\left(\ket{\uparrow\uparrow}+\ket{\downarrow\downarrow}\right)/\sqrt{2}$ and $\ket{t_z}=-i\left(\ket{\uparrow\downarrow}+\ket{\downarrow\uparrow}\right)/\sqrt{2}$. In this basis, the bare spin operators are represented as\cite{BondOperator},
%Because at low temperature the low lying excitations are triplons, we can restrict the full Hilbert space of the spin system into the Hilbert-space of singlet and triplons using bond operator formalism\cite{BondOperator, phase4}, 

\begin{align}
\hat{S}^\alpha_{j,1}&=\frac{i}{2}\left(\hat{t}^{\alpha\dagger}_j \hat{s}_j-\hat{s}^\dagger_j \hat{t}^\alpha_j\right)-\frac{i}{2} \epsilon_{\alpha,\beta,\gamma} \hat{t}^{\beta\dagger}_{j}\hat{t}^\gamma_j, \nonumber\\
\hat{S}^\alpha_{j,2}&=-\frac{i}{2}\left(\hat{t}^{\alpha\dagger}_j \hat{s}_j-\hat{s}^\dagger_j \hat{t}^\alpha_j\right)-\frac{i}{2} \epsilon_{\alpha,\beta,\gamma} \hat{t}^{\beta\dagger}_{j}\hat{t}^\gamma_j,
\end{align}
where $\alpha$, $\beta$ or $\gamma$ represent $x$, $y$ or $z$. The operator  
$\hat{S}^\alpha_{j,k}$ represents the spin operator at site-$k$ (site-1 or site-2) 
of dimer-$j$ (dimer-A or dimer-B), see Fig.\ref{lattice}; $\hat{t}^{\alpha\dagger}_j$ 
and $\hat{s}^\dagger_j$ are the triplon and singlet creation operators at $j$-th dimer. The triplon and singlet operators are bosonic quasiparticles  and obey the following constraint on each dimer-bonds,
\begin{equation}
\hat{s}^\dagger_j\hat{s}_j+\sum_j \hat{t}^{\alpha\dagger}_j\hat{t}^\alpha_j=1
\label{constraint}
\end{equation}
The ground state of the Hamiltonian (\ref{eq1}) is well 
approximated by a condensation of singlets on the dimer bonds. 
In the bond operator formalism, this is implemented as 
%At low temperature we can assume that the singlets are condensed and few triplons are excited. So, at low temperature from Eq.\ref{constrain} we can assume that 
$\left\langle\hat{s}^\dagger_j\right\rangle=\left\langle\hat{s}_j\right\rangle=1$. The lowest excitations are isolated 
triplons on each dimer. Assuming a low density of triplons, 
we can derive a minimal model for the triplons by applying
the bond operator formalism together with the constraint
(\ref{constraint}), and the condition $\left\langle\hat{s}^\dagger_j\right\rangle=\left\langle\hat{s}_j\right\rangle=1$
to the parent Hamiltonian (\ref{eq1}), 

\begin{align}
\pazocal{H}=&J\sum_{\bold{r}_i} \left[\txdagn\txn+\tydagn\tyn+\tzdagn\tzn\right]\nonumber \\
+&ig_zh_z\sum_{\bold{r}_i}\left[\txdagn\tyn-\tydagn\txn\right]\nonumber\\
-&\frac{iD_\perp}{2} \sum_{\bold{r}_i}\sum_{\boldsymbol{\delta}} \left[\tydag{}\txn-\txdag{}\tyn+\text{h.c.}\right],
\label{Hamiltonian}
\end{align}
%\begin{align}
%\pazocal{H}=&J\sum_{\bold{r}_i} \left[\txdagn\txn+\tydagn\tyn+\tzdagn\tzn\right]\nonumber \\
%+&ig_zh_z\sum_{\bold{r}_i}\left[\txdagn\tyn-\tydagn\txn\right]\nonumber\\
%-&\frac{iD_\perp}{2} \sum_{\bold{r}_i} \left[\tydag{x}\txn-\txdag{x}\tyn+\tydagn\tx{x}-\txdagn\ty{x}\right.\nonumber\\
%&\left. \quad\quad\quad-\txdag{y}\tyn+\tydag{y}\txn-\txdagn\ty{y}+\tydagn\tx{y}\right],
%\label{Hamiltonian}
%\end{align}
where we have used a mean field decomposition to keep terms up to bilinear
in the triplon operators. Again $\bold{r}_i$ is the postion vector of the i-th dimer and $\boldsymbol{\delta}_x$ or $\boldsymbol{\delta}_y$ are the vectors denoting the relative positions of the dimers as depicted in Fig.\ref{lattice}(b). The triplons hop on an 
effective square lattice as shown in the figure Fig.\ref{lattice}(b). It is further noted 
that in the Hamiltonian (\ref{Hamiltonian}), the sites corresponding to dimer-A and 
dimer-B are mapped to a equivalent site using the following unitary transformation $t^{x}_{\bold{r}_i,A}=t^{x}_{\bold{r}_i}$, $t^{y}_{\bold{r}_i,A}=t^{y\dagger}_{\bold{r}_i}$, $t^{z}_{\bold{r}_i,A}=t^{x}_{\bold{r}_i}$, $t^{x}_{\bold{r}_i,B}=it^{x}_{\bold{r}_i}$, $t^{y}_{\bold{r}_i,B}=it^{y}_{\bold{r}_i}$, $t^{z}_{\bold{r}_i,B}=t^{z}_{\bold{r}_i}$. Moreover for simplicity we have neglected the pair hopping terms which changes the energy perturbatively to the order of $D_\perp^2$ and hence do not change the band structure\cite{triplon}. 
%In the following section we will discuss the triplon band structure and the presence of $U(1)$ symmetry protected Dirac nodal line.
Since the Hamiltonian (\ref{Hamiltonian}) is translationally invariant, 
it is natural to work in the momentum space. The momentum space triplon-Hamiltonian is given as,
\begin{equation}
\pazocal{H}=\sum_\bk \begin{pmatrix}
\hat{t}^{x\dagger}_\bk \\
\hat{t}^{y\dagger}_\bk \\
\hat{t}^{z\dagger}_\bk
\end{pmatrix}
\begin{pmatrix}
J & -i\gamma(\bk) & 0 \\
i\gamma(\bk) & J & 0\\
0 & 0 & J
\end{pmatrix}
\begin{pmatrix}
\hat{t}^{x}_\bk \\
\hat{t}^{y}_\bk \\
\hat{t}^{z}_\bk
\end{pmatrix},
\label{Matrix}
\end{equation}
where $\gamma(\bk)=-g_z h_z+ D_\perp (\cos(k_x)+\cos(k_y))$. The following canonical transformation of the triplon operators, $\hat{t}^{x\dagger}_\bk=\frac{i}{\sqrt{2}} \left(\hat{t}^{1\dagger}_\bk-\hat{t}^{\bar{1}\dagger}_\bk\right)$, $\hat{t}^{y\dagger}_\bk=\frac{1}{\sqrt{2}} \left(\hat{t}^{1\dagger}_\bk+\hat{t}^{\bar{1}\dagger}_\bk\right)$, $\hat{t}^{z\dagger}_\bk=-i\hat{t}^{0\dagger}_\bk$ diagonalizes the Hamiltonian Eq\ref{Matrix} as,
\begin{equation}
\pazocal{H}=\sum_\bk \begin{pmatrix}
\hat{t}^{\bar{1}\dagger}_\bk \\
\hat{t}^{0\dagger}_\bk \\
\hat{t}^{1\dagger}_\bk
\end{pmatrix}
\begin{pmatrix}
J-\gamma(\bk) & 0 & 0 \\
0 & J & 0\\
0 & 0 & J+\gamma(\bk)
\end{pmatrix}
\begin{pmatrix}
\hat{t}^{\bar{1}}_\bk \\
\hat{t}^{0}_\bk \\
\hat{t}^{1}_\bk
\end{pmatrix},
\label{Hamk}
\end{equation}
where the operators $\hat{t}^{\bar{1}\dagger}_\bk$, $\hat{t}^{0\dagger}_\bk$ and $\hat{t}^{1\dagger}_\bk$ create the states $\ket{t^1}=\ket{\uparrow\uparrow}$, $\ket{t^0}=(\ket{\uparrow\downarrow}+\ket{\downarrow\uparrow})$ and $\ket{t^{\bar{1}}}=\ket{\downarrow\downarrow}$ on the dimer respectively. As a consequence of $U(1)$-symmetry conservation $\hat{S}_z$ is also a eigen-operator of the eigenstates $\ket{t^{\bar{1}}}$, $\ket{t^0}$ and $\ket{t^1}$ and the $S_z$-quantum numbers corresponding to the states  are $-1$, $0$, $+1$ respectively.

%\subsection*{Dirac nodal line}
\section{Results}
\begin{figure}[H]
\includegraphics[width=\textwidth]{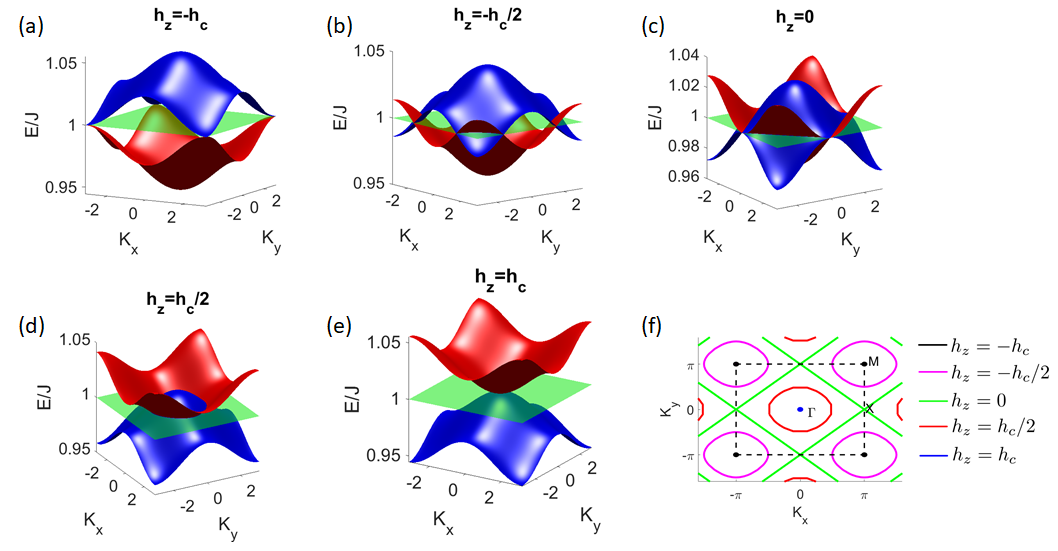} % this command will be ignored
\caption{\textbf{Dirac nodal line magnons.} (a)-(e) The triplon band structure for different magnetic fields. The colors blue, green and red denote the triplon eigenstates $\ket{t^{\bar{1}}}$, $\ket{t^0}$ and $\ket{t^1}$ respectively. For better visibility, the middle band is kept transparent. (f) The k-space position of Dirac nodal line for different magnetic fields.}
\label{NodalLine}
\end{figure}

We have chosen the Hamiltonian parameters as estimated for \ce{SrCu2(BO3)2},
viz., $J=722$ GHz, $D_\perp=-21$ GHz and $g_z=2.28$\cite{triplon, ESR}. 
Using these parameters, we have studied the evolution of the triplon
bands in a longitudinal magnetic field, $h_z$. The band dispersion for 
different magnetic fields are depicted in Fig.\ref{NodalLine}(a)-(e). 
The spectrum consists of one dispersionless, flat band and two
dispersive bands. At zero magnetic field, the two dispersive bands cross
linearly at all points along the perimeter of a quadrilateral formed
by joining the four $X$ points on the Brillouin Zone boundary, shown by
the green lines in Fig.\ref{NodalLine}(f). The dispersion of lower or 
upper bands near nodal line up to second order of $k_{\perp}$ and 
$k_{||}$ is given by, $J\pm D_{\perp} |k_\perp|$, where $k_{||}$ and $k_{\perp}$ denotes the momentum in the direction parallel and perpendicular to the Dirac nodal line respectively. The Dirac nodal line persists 
over a finite range of magnetic field $-h_c<h_z<h_c$, where 
$h_c=\frac{2D_\perp}{g_z}$. Away from $h_z=0$ (and for $|h_z| < h_c$), 
the Dirac nodal line forms a closed loop for any nonzero $h_z$-value 
centered around $\Gamma$ or $M$-point in the Brillouin zone. For  
$sgn(D_zh_z)>0$ ($sgn(D_zh_z)<0$) the closed loop centers around  $\Gamma$($M$) point.  When the magnetic field $h_z$ is close to 
$\pm h_c$ the Dirac nodal line forms a circle with radius 
$k= \cos^{-1}\left(-1+\left|\frac{g_zh_z}{D_\perp}\right|\right)$. The dispersion near the nodal line for the upper or lower bands up to second order of $k_{\perp}$ and 
$k_{||}$ are $J\pm D_{\perp}\sin(k)|k_\perp|\pm\frac{D_\perp}{2} k_{||}^2$, where $k_\perp$ and $k_{||}$ denote the k-points in direction perpendicular and parallel to the circumference of the nodal line.
 %At zero magnetic field the dispersion of lower or upper bands near nodal line up to second order of $k_{\perp}$ and $k_{||}$ is given by, $J\pm D_{\perp} |k_\perp|$. 
At the critical magnetic field $\pm h_c$, the Dirac nodal line shrinks 
to a quadratic band touching point and for magnetic field $|h_z|>|h_c|$ the 
triplon bands become gapped. 

\onecolumngrid

\twocolumngrid

%\subsection*{$U(1)$ symmetry protected topological\\ Dirac nodal line.}
\section{Symmetry considerations}
The Dirac nodal line in the triplon band structure is robust against the perturbation from inversion symmetry breaking term (e.g. the different Heisenberg interaction for dimer-A and dimer-B) as well as time reversal symmetry breaking term(e.g. perpendicular magnetic field). But any perturbation which breaks the U(1)-symmetry lifts the degeneracy at Dirac nodal line. Physically the bond-operator formalism translates the spin model of Shastry-Sutherland lattice into a simple effective square lattice model( Fig.\ref{lattice}(b), Eq.\ref{Hamiltonian}) with three orbitals $\ket{t^{\bar{1}}}$, $\ket{t^0}$ and $\ket{t^1}$ on each lattice site. The presence of $U(1)$-symmetry ensures that the inter-species hopping in the square lattice is disallowed. In other words, the nodal loop exists due to $U(1)$-symmetry. 

We define the following $\mathbb{Z}$-topological invariant for the $U(1)$-symmetry protected Dirac nodal line \cite{symmetry1, symmetry2},
 \begin{equation}
 \Delta S_z=\left\langle\hat{S}^z(\bk_{in})\right\rangle-\left\langle\hat{S}^z(\bk_{out})\right\rangle,
 \end{equation}
where $\bk_{in}$ and $\bk_{out}$ are the k-points inside and outside of the Dirac nodal loop. Again $\left\langle\hat{S}^z(\bk_{l})\right\rangle$ is the eigenvalue of $\hat{S}^z$ at $\bk_l$-point for the bands lower than the plane of Dirac nodal line. For the triplon bands in Fig.\ref{NodalLine}(b)-(d) the topological invariant is $\Delta S_z=+2$. The non-zero topological invariant ensures the robustness of the Dirac nodal line in the presence of additional (perturbative) interactions that do not break the $U(1)$ symmetry.
%under infinitesimal perturbative $U(1)$-preserving term.  

\begin{figure}[H]
\includegraphics[width=0.5\textwidth]{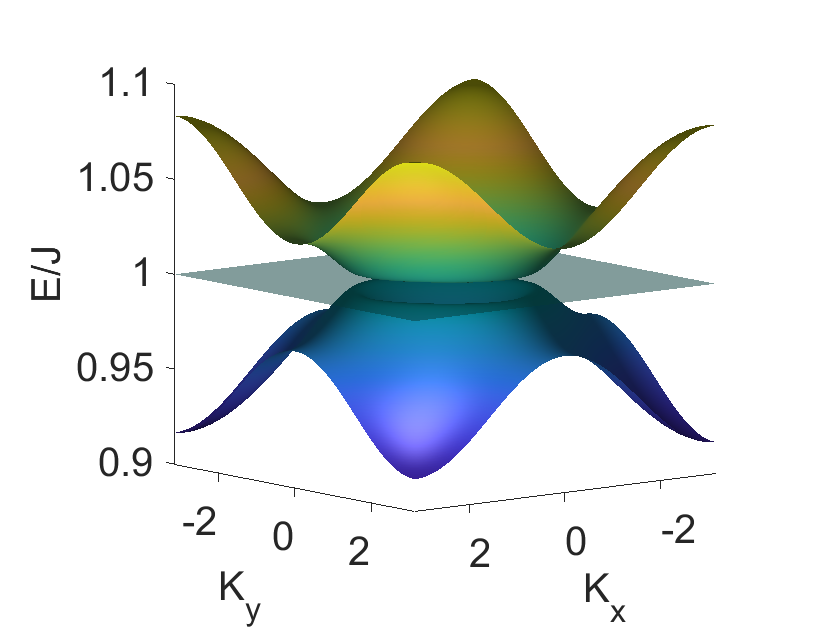} % this command will be ignored
\caption{\textbf{U(1) symmetry breaking.} The triplon band structure for $h_z=h_c/2$, $h_x=h_c/10$. For better visibility the middle band is kept transparent.}
\label{U(1)symmetry}
\end{figure}

To verify that the nodal line is indeed protected by the $U(1)$ symmetry,
we investigate the effects of several perturbations breaking different symmetries of the lattice.
First, we include an in-plane magnetic field $h_x$ along x-direction to illustrate the effect of $U(1)$-symmetry breaking. This transforms the Hamiltonian Eq.\ref{Hamiltonian} as,
\begin{equation}
\pazocal{H}=\sum_\bk \begin{pmatrix}
\hat{t}^{\bar{1}\dagger}_\bk \\
\hat{t}^{0\dagger}_\bk \\
\hat{t}^{1\dagger}_\bk
\end{pmatrix}
\begin{pmatrix}
J-\gamma(\bk) & -\frac{1}{\sqrt{2}}g_xh_x & 0 \\
-\frac{1}{\sqrt{2}}g_xh_x & J & -\frac{1}{\sqrt{2}}g_xh_x\\
0 & -\frac{1}{\sqrt{2}}g_xh_x & J+\gamma(\bk)
\end{pmatrix}
\begin{pmatrix}
\hat{t}^{\bar{1}}_\bk \\
\hat{t}^{0}_\bk \\
\hat{t}^{1}_\bk
\end{pmatrix},
\label{Hamiltonian2}
\end{equation}
where $g_x$ is the g-factor and for simplicity we assume $g_x=g_z$. The eigenvalues of the Hamiltonian Eq.\ref{Hamiltonian2} are $J\pm d(\bk)$, J, where $d(\bk)=(\gamma(\bk)^2+(g_zh_x)^2)^{1/2}$. The triplon bands in presence of in plane magnetic field $h_x$ (Fig.\ref{U(1)symmetry}) shows that the degeneracies at the Dirac nodal line is lifted.

Conservation of time-reversal symmetry along with the U(1)-symmetry further restricts that the nodal line can not be gapped out. In presence of time-reversal symmetry breaking term which preserves the $U(1)$ symmetry, the equivalency between the states $\left|t^{\bar{1}}\right\rangle$ and $\left|t^{1}\right\rangle$ is lifted and so tuning such parameter, the nodal line degeneracy can be lifted. As we can see in figure Fig.\ref{NodalLine}(a),(e) the nodal line degeneracy is lifted tuning the perpendicular magnetic field which preserves the $U(1)$-symmetry but breaks the time reversal symmetry of the system.

\begin{figure}[H]
\includegraphics[width=0.5\textwidth]{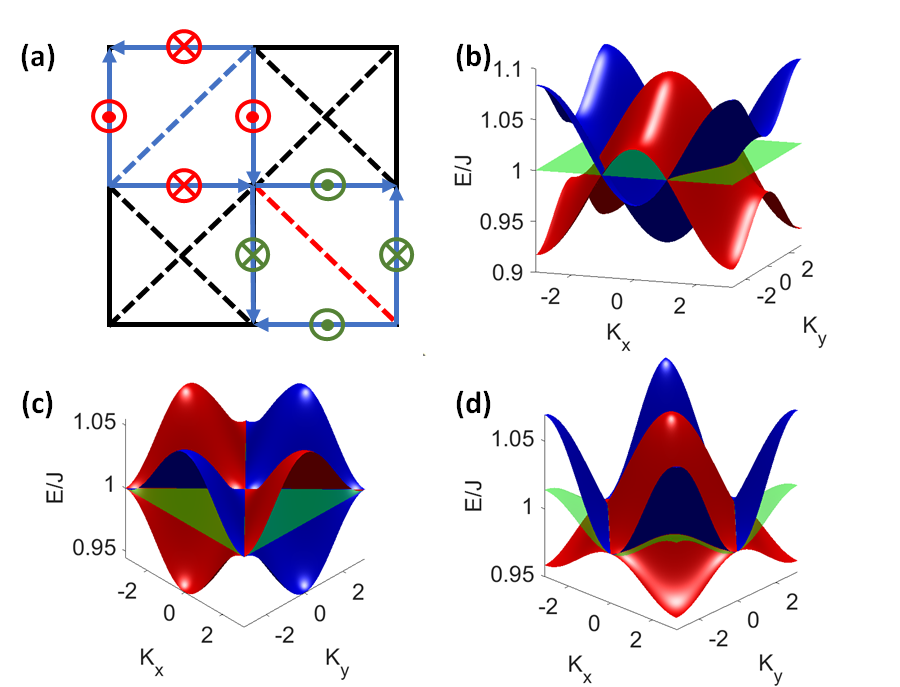} % this command will be ignored
\caption{\textbf{$U(1)\bigotimes \pazocal{T}$ symmetry protected nodal line.} (a) The red and blue circled dots or cross denotes the interaction $D_{\perp,x}$ and $D_{\perp,y}$. The black dotted lines denote the third Heisenberg interaction. The band structure for (b) $D_{\perp,y}=2D_{\perp,x}=-40 GHz$, $J_3=0GHz$; (c) $D_{\perp,y}=-D_{\perp,x}=-20GHz$, $J_3=0GHz$; (d) $D_{\perp,y}=D_{\perp,x}=-20GHz$, $J_3=10GHz$. For all the band structure the other parameters are $J=722GHz$, $g_z=2.28GHz$, $h_z=0$ }
\label{Extra}
\end{figure}

The underlying Shastry-Sutherland lattice and the parent hamiltonian possess additional
symmetries, including $C_4$ rotation about an axis perpendicular to the plane of the 
lattice and passing through  the center of an empty plaquette and a ${\cal G}\otimes {\cal T}$
symmetery, consisting of glide plane and time reversal symmetry operations. 
The choice of perpendicular DM-interactions $D_{\perp,x}$ and $D_{\perp,y}$ as shown in Fig.\ref{Extra}(a), with $D_{\perp,x}\neq D_{\perp,y}$ breaks the ${\cal G}\otimes {\cal T}$ and $C_4$-symmetries. The dispersion
of the triplon bands in the presence of such a symmetry breaking perturbation is 
given by $J-\nu \gamma'(\bk)+J_3\cos(k_x)\cos(k_y)$, where we have also added a third
nearest neighbor Heisenberg exchange interaction $J_3$ (shown by black dotted line in Fig.\ref{Extra}(a)) that does not break any symmetry of the system. Here $\nu=+1,$ $-1,$ $0$ for $\ket{t^{1}}$, $\ket{t^0}$ and $\ket{t^{\bar{1}}}$  respectively and $\gamma'(\bk)=g_z h_z+D_{\perp,x}\cos(k_x)+D_{\perp,y}\cos(k_y)$. The band structures for $D_{\perp,x}\neq D_{\perp,y}$ is shown in Fig.\ref{Extra}(b), (c), where the nodal lines are still protected by $U(1)$ and $ \cal{T}$ symmetry. The Heisenberg-interaction $J_3$ changes the band dispersion of the three different triplon species exactly same way and so the nodal line is not lifted in presence of $J_3$ as shown in Fig.\ref{Extra}(d).

 %To further illustrate the robustness of the nodal line in presence of $U(1)\bigotimes \pazocal{T}$ symmetry, we included more possible interactions in the system as in Fig.\ref{Extra}(a). The perpendicular DM-interactions at the two Dimer plaquette are chosen to be different and named as $D_{\perp,x}$ and $D_{\perp,y}$. When $D_{\perp,x}\neq D_{\perp,y}$ , the glide plane symmetry and $C_4$-symmetry(rotation around the center of non-dimer plaquette) of the Shastry-Sutherland lattice are broken. Moreover third nearest neighbour Heisenberg-exchange interaction $J_3$ shown as black dotted line in Fig.\ref{Extra}(a), is one of the possible spin exchange interaction in the system. Including all the interactions in presence of magnetic field the energies of the triplon species are given by $J-\nu \gamma'(\bk)+J_3\cos(k_x)\cos(k_y)$, where $\nu=+1,$ $-1,$ $0$ for $\ket{t^{1}}$, $\ket{t^0}$ and $\ket{t^{\bar{1}}}$ triplon species respectively and $\gamma'(\bk)=g_z h_z+D_{\perp,x}\cos(k_x)+D_{\perp,y}\cos(k_y)$. The band structures for $D_{\perp,x}\neq D_{\perp,y}$ case is shown in Fig.\ref{Extra}(b), (c), where the nodal lines are still protected by $U(1)\bigotimes \pazocal{T}$ symmetry. Moreover, The Heisenberg-interaction $J_3$ changes the band dispersion of the three different triplon species exactly same way and so the nodal lines is not lifted in presence of $J_3$ as shown in Fig.\ref{Extra}(d).

\section{Conclusion}
We have shown that triplon Dirac nodal line appears in the magnetic excitation
spectrum of a microscopic model of \ce{SrCu2(BO3)2} in its high symmetry phase using experimentally
determined hamiltonian parameters. Whereas the previous studies showed the presence of nodal lines in the excitation spectrum of 
%a long range ordered ferromagnetic or anti-ferromagnetic ground state, 
{\color{red}classical spin states}, this study shows that the Dirac nodal lines also exist in the excitations of the short-range ordered purely quantum dimerized ground state of Shastry-Sutherland lattice. Our results demonstrate that the nodal lines are protected by 
a $U(1)$ and time reversal symmetries and is robust against any perturbation that do not break 
both of them. The nodal lines also persist in the presence of weak time reversal symmetry breaking fields. In an applied longitudinal magnetic field, the nodal lines continually shrink
and eventually disappear at a critical field. In contrast to previous studies that explore
prototypical models of certain classes of quantum magnets, our results are based on a faithful
microscopic model of a (quasi 2D) real magnet. 
%Furthermore, our results establish the appearance of
%triplon nodel lines in 2d magnets, without the need for any additional inter-layer coupling, beyond of a two-dimensional magnetic system which is essentially bosonic analogue to the fermionic counterpart of Dirac nodal loop in two-dimensional crystals\cite{2DSemimetal,2DSemimetal2,2DSemimetal3,2DSemimetal4}. In contrast to the bilayer honeycomb ferromagnet\cite{Quasi2D}, the magnetic system considered in this study is perfectly two-dimensional and no interlayer coupling is necessary. 
This novel Dirac nodal loop can be observed via inelastic neutron scattering experiment in the high crystal symmetry phase of \ce{SrCu2(BO3)2}, when all the in-plane DM-interactions are disallowed by symmetry of the crystal geometry thus protecting the $U(1)$-symmetry of the spin-system.
%The high symmetry phase of \ce{SrCu2(BO3)2} exists above a temperature 395K\cite{structure2}, we propose at low temperature a pressure induced structural phase transition can be induced to set all the magnetic moment carrying Cu$^{2+}$-ion exists on the same plane. Again, 
The presence of $U(1)$-symmetry assures the presence of the Dirac nodal line and so it is expected that at higher temperature the triplon-triplon interactions (which has not been considered in our study) can not remove the Dirac-nodal line\cite{DiracMagnon,DiracMagnon2,DiracMagnon3}.

Financial support from the Ministry of Education, Singapore, in the form
of grant MOE2018-T1-001-021 is gratefully acknowledged.
\pagebreak
\appendix

\bibliographystyle{apsrev4-1}

\begin{thebibliography}{75}%
\makeatletter
\providecommand \@ifxundefined [1]{%
 \@ifx{#1\undefined}
}%
\providecommand \@ifnum [1]{%
 \ifnum #1\expandafter \@firstoftwo
 \else \expandafter \@secondoftwo
 \fi
}%
\providecommand \@ifx [1]{%
 \ifx #1\expandafter \@firstoftwo
 \else \expandafter \@secondoftwo
 \fi
}%
\providecommand \natexlab [1]{#1}%
\providecommand \enquote  [1]{``#1''}%
\providecommand \bibnamefont  [1]{#1}%
\providecommand \bibfnamefont [1]{#1}%
\providecommand \citenamefont [1]{#1}%
\providecommand \href@noop [0]{\@secondoftwo}%
\providecommand \href [0]{\begingroup \@sanitize@url \@href}%
\providecommand \@href[1]{\@@startlink{#1}\@@href}%
\providecommand \@@href[1]{\endgroup#1\@@endlink}%
\providecommand \@sanitize@url [0]{\catcode `\\12\catcode `\$12\catcode
  `\&12\catcode `\#12\catcode `\^12\catcode `\_12\catcode `\%12\relax}%
\providecommand \@@startlink[1]{}%
\providecommand \@@endlink[0]{}%
\providecommand \url  [0]{\begingroup\@sanitize@url \@url }%
\providecommand \@url [1]{\endgroup\@href {#1}{\urlprefix }}%
\providecommand \urlprefix  [0]{URL }%
\providecommand \Eprint [0]{\href }%
\providecommand \doibase [0]{http://dx.doi.org/}%
\providecommand \selectlanguage [0]{\@gobble}%
\providecommand \bibinfo  [0]{\@secondoftwo}%
\providecommand \bibfield  [0]{\@secondoftwo}%
\providecommand \translation [1]{[#1]}%
\providecommand \BibitemOpen [0]{}%
\providecommand \bibitemStop [0]{}%
\providecommand \bibitemNoStop [0]{.\EOS\space}%
\providecommand \EOS [0]{\spacefactor3000\relax}%
\providecommand \BibitemShut  [1]{\csname bibitem#1\endcsname}%
\let\auto@bib@innerbib\@empty
%</preamble>
\bibitem [{\citenamefont {Lv}\ \emph {et~al.}(2015)\citenamefont {Lv},
  \citenamefont {Weng}, \citenamefont {Fu}, \citenamefont {Wang}, \citenamefont
  {Miao}, \citenamefont {Ma}, \citenamefont {Richard}, \citenamefont {Huang},
  \citenamefont {Zhao}, \citenamefont {Chen}, \citenamefont {Fang},
  \citenamefont {Dai}, \citenamefont {Qian},\ and\ \citenamefont
  {Ding}}]{WeylSemiMetal}%
  \BibitemOpen
  \bibfield  {author} {\bibinfo {author} {\bibfnamefont {B.~Q.}\ \bibnamefont
  {Lv}}, \bibinfo {author} {\bibfnamefont {H.~M.}\ \bibnamefont {Weng}},
  \bibinfo {author} {\bibfnamefont {B.~B.}\ \bibnamefont {Fu}}, \bibinfo
  {author} {\bibfnamefont {X.~P.}\ \bibnamefont {Wang}}, \bibinfo {author}
  {\bibfnamefont {H.}~\bibnamefont {Miao}}, \bibinfo {author} {\bibfnamefont
  {J.}~\bibnamefont {Ma}}, \bibinfo {author} {\bibfnamefont {P.}~\bibnamefont
  {Richard}}, \bibinfo {author} {\bibfnamefont {X.~C.}\ \bibnamefont {Huang}},
  \bibinfo {author} {\bibfnamefont {L.~X.}\ \bibnamefont {Zhao}}, \bibinfo
  {author} {\bibfnamefont {G.~F.}\ \bibnamefont {Chen}}, \bibinfo {author}
  {\bibfnamefont {Z.}~\bibnamefont {Fang}}, \bibinfo {author} {\bibfnamefont
  {X.}~\bibnamefont {Dai}}, \bibinfo {author} {\bibfnamefont {T.}~\bibnamefont
  {Qian}}, \ and\ \bibinfo {author} {\bibfnamefont {H.}~\bibnamefont {Ding}},\
  }\href {\doibase 10.1103/PhysRevX.5.031013} {\bibfield  {journal} {\bibinfo
  {journal} {Phys. Rev. X}\ }\textbf {\bibinfo {volume} {5}},\ \bibinfo {pages}
  {031013} (\bibinfo {year} {2015})}\BibitemShut {NoStop}%
\bibitem [{\citenamefont {Weng}\ \emph {et~al.}(2015)\citenamefont {Weng},
  \citenamefont {Fang}, \citenamefont {Fang}, \citenamefont {Bernevig},\ and\
  \citenamefont {Dai}}]{WeylSemiMetal2}%
  \BibitemOpen
  \bibfield  {author} {\bibinfo {author} {\bibfnamefont {H.}~\bibnamefont
  {Weng}}, \bibinfo {author} {\bibfnamefont {C.}~\bibnamefont {Fang}}, \bibinfo
  {author} {\bibfnamefont {Z.}~\bibnamefont {Fang}}, \bibinfo {author}
  {\bibfnamefont {B.~A.}\ \bibnamefont {Bernevig}}, \ and\ \bibinfo {author}
  {\bibfnamefont {X.}~\bibnamefont {Dai}},\ }\href {\doibase
  10.1103/PhysRevX.5.011029} {\bibfield  {journal} {\bibinfo  {journal} {Phys.
  Rev. X}\ }\textbf {\bibinfo {volume} {5}},\ \bibinfo {pages} {011029}
  (\bibinfo {year} {2015})}\BibitemShut {NoStop}%
\bibitem [{\citenamefont {Huang}\ \emph {et~al.}(2015)\citenamefont {Huang},
  \citenamefont {Xu}, \citenamefont {Belopolski}, \citenamefont {Lee},
  \citenamefont {Chang}, \citenamefont {Wang}, \citenamefont {Alidoust},
  \citenamefont {Bian}, \citenamefont {Neupane}, \citenamefont {Zhang},
  \citenamefont {Jia}, \citenamefont {Bansil}, \citenamefont {Lin},\ and\
  \citenamefont {Hasan}}]{WeylSemiMetal3}%
  \BibitemOpen
  \bibfield  {author} {\bibinfo {author} {\bibfnamefont {S.-M.}\ \bibnamefont
  {Huang}}, \bibinfo {author} {\bibfnamefont {S.-Y.}\ \bibnamefont {Xu}},
  \bibinfo {author} {\bibfnamefont {I.}~\bibnamefont {Belopolski}}, \bibinfo
  {author} {\bibfnamefont {C.-C.}\ \bibnamefont {Lee}}, \bibinfo {author}
  {\bibfnamefont {G.}~\bibnamefont {Chang}}, \bibinfo {author} {\bibfnamefont
  {B.}~\bibnamefont {Wang}}, \bibinfo {author} {\bibfnamefont {N.}~\bibnamefont
  {Alidoust}}, \bibinfo {author} {\bibfnamefont {G.}~\bibnamefont {Bian}},
  \bibinfo {author} {\bibfnamefont {M.}~\bibnamefont {Neupane}}, \bibinfo
  {author} {\bibfnamefont {C.}~\bibnamefont {Zhang}}, \bibinfo {author}
  {\bibfnamefont {S.}~\bibnamefont {Jia}}, \bibinfo {author} {\bibfnamefont
  {A.}~\bibnamefont {Bansil}}, \bibinfo {author} {\bibfnamefont
  {H.}~\bibnamefont {Lin}}, \ and\ \bibinfo {author} {\bibfnamefont {M.~Z.}\
  \bibnamefont {Hasan}},\ }\href {\doibase 10.1038/ncomms8373} {\bibfield
  {journal} {\bibinfo  {journal} {Nature Communications}\ }\textbf {\bibinfo
  {volume} {6}},\ \bibinfo {pages} {7373} (\bibinfo {year} {2015})}\BibitemShut
  {NoStop}%
\bibitem [{\citenamefont {Burkov}\ \emph {et~al.}(2011)\citenamefont {Burkov},
  \citenamefont {Hook},\ and\ \citenamefont {Balents}}]{NL1}%
  \BibitemOpen
  \bibfield  {author} {\bibinfo {author} {\bibfnamefont {A.~A.}\ \bibnamefont
  {Burkov}}, \bibinfo {author} {\bibfnamefont {M.~D.}\ \bibnamefont {Hook}}, \
  and\ \bibinfo {author} {\bibfnamefont {L.}~\bibnamefont {Balents}},\ }\href
  {\doibase 10.1103/PhysRevB.84.235126} {\bibfield  {journal} {\bibinfo
  {journal} {Phys. Rev. B}\ }\textbf {\bibinfo {volume} {84}},\ \bibinfo
  {pages} {235126} (\bibinfo {year} {2011})}\BibitemShut {NoStop}%
\bibitem [{\citenamefont {Xu}\ \emph {et~al.}(2011)\citenamefont {Xu},
  \citenamefont {Weng}, \citenamefont {Wang}, \citenamefont {Dai},\ and\
  \citenamefont {Fang}}]{NL2}%
  \BibitemOpen
  \bibfield  {author} {\bibinfo {author} {\bibfnamefont {G.}~\bibnamefont
  {Xu}}, \bibinfo {author} {\bibfnamefont {H.}~\bibnamefont {Weng}}, \bibinfo
  {author} {\bibfnamefont {Z.}~\bibnamefont {Wang}}, \bibinfo {author}
  {\bibfnamefont {X.}~\bibnamefont {Dai}}, \ and\ \bibinfo {author}
  {\bibfnamefont {Z.}~\bibnamefont {Fang}},\ }\href {\doibase
  10.1103/PhysRevLett.107.186806} {\bibfield  {journal} {\bibinfo  {journal}
  {Phys. Rev. Lett.}\ }\textbf {\bibinfo {volume} {107}},\ \bibinfo {pages}
  {186806} (\bibinfo {year} {2011})}\BibitemShut {NoStop}%
\bibitem [{\citenamefont {Fang}\ \emph {et~al.}(2015)\citenamefont {Fang},
  \citenamefont {Chen}, \citenamefont {Kee},\ and\ \citenamefont {Fu}}]{NL3}%
  \BibitemOpen
  \bibfield  {author} {\bibinfo {author} {\bibfnamefont {C.}~\bibnamefont
  {Fang}}, \bibinfo {author} {\bibfnamefont {Y.}~\bibnamefont {Chen}}, \bibinfo
  {author} {\bibfnamefont {H.-Y.}\ \bibnamefont {Kee}}, \ and\ \bibinfo
  {author} {\bibfnamefont {L.}~\bibnamefont {Fu}},\ }\href {\doibase
  10.1103/PhysRevB.92.081201} {\bibfield  {journal} {\bibinfo  {journal} {Phys.
  Rev. B}\ }\textbf {\bibinfo {volume} {92}},\ \bibinfo {pages} {081201}
  (\bibinfo {year} {2015})}\BibitemShut {NoStop}%
\bibitem [{\citenamefont {Kim}\ \emph {et~al.}(2015)\citenamefont {Kim},
  \citenamefont {Wieder}, \citenamefont {Kane},\ and\ \citenamefont
  {Rappe}}]{NL4}%
  \BibitemOpen
  \bibfield  {author} {\bibinfo {author} {\bibfnamefont {Y.}~\bibnamefont
  {Kim}}, \bibinfo {author} {\bibfnamefont {B.~J.}\ \bibnamefont {Wieder}},
  \bibinfo {author} {\bibfnamefont {C.~L.}\ \bibnamefont {Kane}}, \ and\
  \bibinfo {author} {\bibfnamefont {A.~M.}\ \bibnamefont {Rappe}},\ }\href
  {\doibase 10.1103/PhysRevLett.115.036806} {\bibfield  {journal} {\bibinfo
  {journal} {Phys. Rev. Lett.}\ }\textbf {\bibinfo {volume} {115}},\ \bibinfo
  {pages} {036806} (\bibinfo {year} {2015})}\BibitemShut {NoStop}%
\bibitem [{\citenamefont {Yu}\ \emph {et~al.}(2015)\citenamefont {Yu},
  \citenamefont {Weng}, \citenamefont {Fang}, \citenamefont {Dai},\ and\
  \citenamefont {Hu}}]{NL5}%
  \BibitemOpen
  \bibfield  {author} {\bibinfo {author} {\bibfnamefont {R.}~\bibnamefont
  {Yu}}, \bibinfo {author} {\bibfnamefont {H.}~\bibnamefont {Weng}}, \bibinfo
  {author} {\bibfnamefont {Z.}~\bibnamefont {Fang}}, \bibinfo {author}
  {\bibfnamefont {X.}~\bibnamefont {Dai}}, \ and\ \bibinfo {author}
  {\bibfnamefont {X.}~\bibnamefont {Hu}},\ }\href {\doibase
  10.1103/PhysRevLett.115.036807} {\bibfield  {journal} {\bibinfo  {journal}
  {Phys. Rev. Lett.}\ }\textbf {\bibinfo {volume} {115}},\ \bibinfo {pages}
  {036807} (\bibinfo {year} {2015})}\BibitemShut {NoStop}%
\bibitem [{\citenamefont {Huh}\ \emph {et~al.}(2016)\citenamefont {Huh},
  \citenamefont {Moon},\ and\ \citenamefont {Kim}}]{NL6}%
  \BibitemOpen
  \bibfield  {author} {\bibinfo {author} {\bibfnamefont {Y.}~\bibnamefont
  {Huh}}, \bibinfo {author} {\bibfnamefont {E.-G.}\ \bibnamefont {Moon}}, \
  and\ \bibinfo {author} {\bibfnamefont {Y.~B.}\ \bibnamefont {Kim}},\ }\href
  {\doibase 10.1103/PhysRevB.93.035138} {\bibfield  {journal} {\bibinfo
  {journal} {Phys. Rev. B}\ }\textbf {\bibinfo {volume} {93}},\ \bibinfo
  {pages} {035138} (\bibinfo {year} {2016})}\BibitemShut {NoStop}%
\bibitem [{\citenamefont {Rhim}\ and\ \citenamefont {Kim}(2015)}]{NL7}%
  \BibitemOpen
  \bibfield  {author} {\bibinfo {author} {\bibfnamefont {J.-W.}\ \bibnamefont
  {Rhim}}\ and\ \bibinfo {author} {\bibfnamefont {Y.~B.}\ \bibnamefont {Kim}},\
  }\href {\doibase 10.1103/PhysRevB.92.045126} {\bibfield  {journal} {\bibinfo
  {journal} {Phys. Rev. B}\ }\textbf {\bibinfo {volume} {92}},\ \bibinfo
  {pages} {045126} (\bibinfo {year} {2015})}\BibitemShut {NoStop}%
\bibitem [{\citenamefont {{Wu}}\ \emph {et~al.}(2016)\citenamefont {{Wu}},
  \citenamefont {{Wang}}, \citenamefont {{Mun}}, \citenamefont {{Johnson}},
  \citenamefont {{Mou}}, \citenamefont {{Huang}}, \citenamefont {{Lee}},
  \citenamefont {{Bud'Ko}}, \citenamefont {{Canfield}},\ and\ \citenamefont
  {{Kaminski}}}]{NL8}%
  \BibitemOpen
  \bibfield  {author} {\bibinfo {author} {\bibfnamefont {Y.}~\bibnamefont
  {{Wu}}}, \bibinfo {author} {\bibfnamefont {L.-L.}\ \bibnamefont {{Wang}}},
  \bibinfo {author} {\bibfnamefont {E.}~\bibnamefont {{Mun}}}, \bibinfo
  {author} {\bibfnamefont {D.~D.}\ \bibnamefont {{Johnson}}}, \bibinfo {author}
  {\bibfnamefont {D.}~\bibnamefont {{Mou}}}, \bibinfo {author} {\bibfnamefont
  {L.}~\bibnamefont {{Huang}}}, \bibinfo {author} {\bibfnamefont
  {Y.}~\bibnamefont {{Lee}}}, \bibinfo {author} {\bibfnamefont {S.~L.}\
  \bibnamefont {{Bud'Ko}}}, \bibinfo {author} {\bibfnamefont {P.~C.}\
  \bibnamefont {{Canfield}}}, \ and\ \bibinfo {author} {\bibfnamefont
  {A.}~\bibnamefont {{Kaminski}}},\ }\href {\doibase 10.1038/nphys3712}
  {\bibfield  {journal} {\bibinfo  {journal} {Nature Physics}\ }\textbf
  {\bibinfo {volume} {12}},\ \bibinfo {pages} {667} (\bibinfo {year} {2016})},\
  \Eprint {http://arxiv.org/abs/1603.00934} {arXiv:1603.00934
  [cond-mat.mtrl-sci]} \BibitemShut {NoStop}%
\bibitem [{\citenamefont {Bian}\ \emph
  {et~al.}(2016{\natexlab{a}})\citenamefont {Bian}, \citenamefont {Chang},
  \citenamefont {Sankar}, \citenamefont {Xu}, \citenamefont {Zheng},
  \citenamefont {Neupert}, \citenamefont {Chiu}, \citenamefont {Huang},
  \citenamefont {Chang}, \citenamefont {Belopolski}, \citenamefont {Sanchez},
  \citenamefont {Neupane}, \citenamefont {Alidoust}, \citenamefont {Liu},
  \citenamefont {Wang}, \citenamefont {Lee}, \citenamefont {Jeng},
  \citenamefont {Zhang}, \citenamefont {Yuan},\ and\ \citenamefont
  {Hasan}}]{Mirror3}%
  \BibitemOpen
  \bibfield  {author} {\bibinfo {author} {\bibfnamefont {G.}~\bibnamefont
  {Bian}}, \bibinfo {author} {\bibfnamefont {T.-R.}\ \bibnamefont {Chang}},
  \bibinfo {author} {\bibfnamefont {R.}~\bibnamefont {Sankar}}, \bibinfo
  {author} {\bibfnamefont {S.-Y.}\ \bibnamefont {Xu}}, \bibinfo {author}
  {\bibfnamefont {H.}~\bibnamefont {Zheng}}, \bibinfo {author} {\bibfnamefont
  {T.}~\bibnamefont {Neupert}}, \bibinfo {author} {\bibfnamefont {C.-K.}\
  \bibnamefont {Chiu}}, \bibinfo {author} {\bibfnamefont {S.-M.}\ \bibnamefont
  {Huang}}, \bibinfo {author} {\bibfnamefont {G.}~\bibnamefont {Chang}},
  \bibinfo {author} {\bibfnamefont {I.}~\bibnamefont {Belopolski}}, \bibinfo
  {author} {\bibfnamefont {D.}~\bibnamefont {Sanchez}}, \bibinfo {author}
  {\bibfnamefont {M.}~\bibnamefont {Neupane}}, \bibinfo {author} {\bibfnamefont
  {N.}~\bibnamefont {Alidoust}}, \bibinfo {author} {\bibfnamefont
  {C.}~\bibnamefont {Liu}}, \bibinfo {author} {\bibfnamefont {B.}~\bibnamefont
  {Wang}}, \bibinfo {author} {\bibfnamefont {C.-C.}\ \bibnamefont {Lee}},
  \bibinfo {author} {\bibfnamefont {H.-T.}\ \bibnamefont {Jeng}}, \bibinfo
  {author} {\bibfnamefont {C.}~\bibnamefont {Zhang}}, \bibinfo {author}
  {\bibfnamefont {Z.}~\bibnamefont {Yuan}}, \ and\ \bibinfo {author}
  {\bibfnamefont {M.~Z.}\ \bibnamefont {Hasan}},\ }\href {\doibase
  10.1038/ncomms10556} {\bibfield  {journal} {\bibinfo  {journal} {Nature
  Communications}\ }\textbf {\bibinfo {volume} {7}},\ \bibinfo {pages} {10556}
  (\bibinfo {year} {2016}{\natexlab{a}})}\BibitemShut {NoStop}%
\bibitem [{\citenamefont {Bian}\ \emph
  {et~al.}(2016{\natexlab{b}})\citenamefont {Bian}, \citenamefont {Chang},
  \citenamefont {Zheng}, \citenamefont {Velury}, \citenamefont {Xu},
  \citenamefont {Neupert}, \citenamefont {Chiu}, \citenamefont {Huang},
  \citenamefont {Sanchez}, \citenamefont {Belopolski}, \citenamefont
  {Alidoust}, \citenamefont {Chen}, \citenamefont {Chang}, \citenamefont
  {Bansil}, \citenamefont {Jeng}, \citenamefont {Lin},\ and\ \citenamefont
  {Hasan}}]{NL10}%
  \BibitemOpen
  \bibfield  {author} {\bibinfo {author} {\bibfnamefont {G.}~\bibnamefont
  {Bian}}, \bibinfo {author} {\bibfnamefont {T.-R.}\ \bibnamefont {Chang}},
  \bibinfo {author} {\bibfnamefont {H.}~\bibnamefont {Zheng}}, \bibinfo
  {author} {\bibfnamefont {S.}~\bibnamefont {Velury}}, \bibinfo {author}
  {\bibfnamefont {S.-Y.}\ \bibnamefont {Xu}}, \bibinfo {author} {\bibfnamefont
  {T.}~\bibnamefont {Neupert}}, \bibinfo {author} {\bibfnamefont {C.-K.}\
  \bibnamefont {Chiu}}, \bibinfo {author} {\bibfnamefont {S.-M.}\ \bibnamefont
  {Huang}}, \bibinfo {author} {\bibfnamefont {D.~S.}\ \bibnamefont {Sanchez}},
  \bibinfo {author} {\bibfnamefont {I.}~\bibnamefont {Belopolski}}, \bibinfo
  {author} {\bibfnamefont {N.}~\bibnamefont {Alidoust}}, \bibinfo {author}
  {\bibfnamefont {P.-J.}\ \bibnamefont {Chen}}, \bibinfo {author}
  {\bibfnamefont {G.}~\bibnamefont {Chang}}, \bibinfo {author} {\bibfnamefont
  {A.}~\bibnamefont {Bansil}}, \bibinfo {author} {\bibfnamefont {H.-T.}\
  \bibnamefont {Jeng}}, \bibinfo {author} {\bibfnamefont {H.}~\bibnamefont
  {Lin}}, \ and\ \bibinfo {author} {\bibfnamefont {M.~Z.}\ \bibnamefont
  {Hasan}},\ }\href {\doibase 10.1103/PhysRevB.93.121113} {\bibfield  {journal}
  {\bibinfo  {journal} {Phys. Rev. B}\ }\textbf {\bibinfo {volume} {93}},\
  \bibinfo {pages} {121113} (\bibinfo {year} {2016}{\natexlab{b}})}\BibitemShut
  {NoStop}%
\bibitem [{\citenamefont {Schoop}\ \emph {et~al.}(2016)\citenamefont {Schoop},
  \citenamefont {Ali}, \citenamefont {Straßer}, \citenamefont {Topp},
  \citenamefont {Varykhalov}, \citenamefont {Marchenko}, \citenamefont
  {Duppel}, \citenamefont {Parkin}, \citenamefont {Lotsch},\ and\ \citenamefont
  {Ast}}]{NL11}%
  \BibitemOpen
  \bibfield  {author} {\bibinfo {author} {\bibfnamefont {L.~M.}\ \bibnamefont
  {Schoop}}, \bibinfo {author} {\bibfnamefont {M.~N.}\ \bibnamefont {Ali}},
  \bibinfo {author} {\bibfnamefont {C.}~\bibnamefont {Straßer}}, \bibinfo
  {author} {\bibfnamefont {A.}~\bibnamefont {Topp}}, \bibinfo {author}
  {\bibfnamefont {A.}~\bibnamefont {Varykhalov}}, \bibinfo {author}
  {\bibfnamefont {D.}~\bibnamefont {Marchenko}}, \bibinfo {author}
  {\bibfnamefont {V.}~\bibnamefont {Duppel}}, \bibinfo {author} {\bibfnamefont
  {S.~S.~P.}\ \bibnamefont {Parkin}}, \bibinfo {author} {\bibfnamefont {B.~V.}\
  \bibnamefont {Lotsch}}, \ and\ \bibinfo {author} {\bibfnamefont {C.~R.}\
  \bibnamefont {Ast}},\ }\href {\doibase 10.1038/ncomms11696} {\bibfield
  {journal} {\bibinfo  {journal} {Nature Communications}\ }\textbf {\bibinfo
  {volume} {7}} (\bibinfo {year} {2016}),\ 10.1038/ncomms11696}\BibitemShut
  {NoStop}%
\bibitem [{\citenamefont {Neupane}\ \emph {et~al.}(2016)\citenamefont
  {Neupane}, \citenamefont {Belopolski}, \citenamefont {Hosen}, \citenamefont
  {Sanchez}, \citenamefont {Sankar}, \citenamefont {Szlawska}, \citenamefont
  {Xu}, \citenamefont {Dimitri}, \citenamefont {Dhakal}, \citenamefont
  {Maldonado}, \citenamefont {Oppeneer}, \citenamefont {Kaczorowski},
  \citenamefont {Chou}, \citenamefont {Hasan},\ and\ \citenamefont
  {Durakiewicz}}]{NL12}%
  \BibitemOpen
  \bibfield  {author} {\bibinfo {author} {\bibfnamefont {M.}~\bibnamefont
  {Neupane}}, \bibinfo {author} {\bibfnamefont {I.}~\bibnamefont {Belopolski}},
  \bibinfo {author} {\bibfnamefont {M.~M.}\ \bibnamefont {Hosen}}, \bibinfo
  {author} {\bibfnamefont {D.~S.}\ \bibnamefont {Sanchez}}, \bibinfo {author}
  {\bibfnamefont {R.}~\bibnamefont {Sankar}}, \bibinfo {author} {\bibfnamefont
  {M.}~\bibnamefont {Szlawska}}, \bibinfo {author} {\bibfnamefont {S.-Y.}\
  \bibnamefont {Xu}}, \bibinfo {author} {\bibfnamefont {K.}~\bibnamefont
  {Dimitri}}, \bibinfo {author} {\bibfnamefont {N.}~\bibnamefont {Dhakal}},
  \bibinfo {author} {\bibfnamefont {P.}~\bibnamefont {Maldonado}}, \bibinfo
  {author} {\bibfnamefont {P.~M.}\ \bibnamefont {Oppeneer}}, \bibinfo {author}
  {\bibfnamefont {D.}~\bibnamefont {Kaczorowski}}, \bibinfo {author}
  {\bibfnamefont {F.}~\bibnamefont {Chou}}, \bibinfo {author} {\bibfnamefont
  {M.~Z.}\ \bibnamefont {Hasan}}, \ and\ \bibinfo {author} {\bibfnamefont
  {T.}~\bibnamefont {Durakiewicz}},\ }\href {\doibase
  10.1103/PhysRevB.93.201104} {\bibfield  {journal} {\bibinfo  {journal} {Phys.
  Rev. B}\ }\textbf {\bibinfo {volume} {93}},\ \bibinfo {pages} {201104}
  (\bibinfo {year} {2016})}\BibitemShut {NoStop}%
\bibitem [{\citenamefont {Gibson}\ \emph {et~al.}(2015)\citenamefont {Gibson},
  \citenamefont {Schoop}, \citenamefont {Muechler}, \citenamefont {Xie},
  \citenamefont {Hirschberger}, \citenamefont {Ong}, \citenamefont {Car},\ and\
  \citenamefont {Cava}}]{Inversion2}%
  \BibitemOpen
  \bibfield  {author} {\bibinfo {author} {\bibfnamefont {Q.~D.}\ \bibnamefont
  {Gibson}}, \bibinfo {author} {\bibfnamefont {L.~M.}\ \bibnamefont {Schoop}},
  \bibinfo {author} {\bibfnamefont {L.}~\bibnamefont {Muechler}}, \bibinfo
  {author} {\bibfnamefont {L.~S.}\ \bibnamefont {Xie}}, \bibinfo {author}
  {\bibfnamefont {M.}~\bibnamefont {Hirschberger}}, \bibinfo {author}
  {\bibfnamefont {N.~P.}\ \bibnamefont {Ong}}, \bibinfo {author} {\bibfnamefont
  {R.}~\bibnamefont {Car}}, \ and\ \bibinfo {author} {\bibfnamefont {R.~J.}\
  \bibnamefont {Cava}},\ }\href {\doibase 10.1103/PhysRevB.91.205128}
  {\bibfield  {journal} {\bibinfo  {journal} {Phys. Rev. B}\ }\textbf {\bibinfo
  {volume} {91}},\ \bibinfo {pages} {205128} (\bibinfo {year}
  {2015})}\BibitemShut {NoStop}%
\bibitem [{\citenamefont {Yang}\ \emph {et~al.}(2018)\citenamefont {Yang},
  \citenamefont {Yang}, \citenamefont {Derunova}, \citenamefont {Parkin},
  \citenamefont {Yan},\ and\ \citenamefont {Ali}}]{SymmetryDemanded}%
  \BibitemOpen
  \bibfield  {author} {\bibinfo {author} {\bibfnamefont {S.-Y.}\ \bibnamefont
  {Yang}}, \bibinfo {author} {\bibfnamefont {H.}~\bibnamefont {Yang}}, \bibinfo
  {author} {\bibfnamefont {E.}~\bibnamefont {Derunova}}, \bibinfo {author}
  {\bibfnamefont {S.~S.~P.}\ \bibnamefont {Parkin}}, \bibinfo {author}
  {\bibfnamefont {B.}~\bibnamefont {Yan}}, \ and\ \bibinfo {author}
  {\bibfnamefont {M.~N.}\ \bibnamefont {Ali}},\ }\href {\doibase
  10.1080/23746149.2017.1414631} {\bibfield  {journal} {\bibinfo  {journal}
  {Advances in Physics: X}\ }\textbf {\bibinfo {volume} {3}},\ \bibinfo {pages}
  {1414631} (\bibinfo {year} {2018})},\ \Eprint
  {http://arxiv.org/abs/https://doi.org/10.1080/23746149.2017.1414631}
  {https://doi.org/10.1080/23746149.2017.1414631} \BibitemShut {NoStop}%
\bibitem [{\citenamefont {Chan}\ \emph {et~al.}(2016)\citenamefont {Chan},
  \citenamefont {Chiu}, \citenamefont {Chou},\ and\ \citenamefont
  {Schnyder}}]{Mirror2}%
  \BibitemOpen
  \bibfield  {author} {\bibinfo {author} {\bibfnamefont {Y.-H.}\ \bibnamefont
  {Chan}}, \bibinfo {author} {\bibfnamefont {C.-K.}\ \bibnamefont {Chiu}},
  \bibinfo {author} {\bibfnamefont {M.~Y.}\ \bibnamefont {Chou}}, \ and\
  \bibinfo {author} {\bibfnamefont {A.~P.}\ \bibnamefont {Schnyder}},\ }\href
  {\doibase 10.1103/PhysRevB.93.205132} {\bibfield  {journal} {\bibinfo
  {journal} {Phys. Rev. B}\ }\textbf {\bibinfo {volume} {93}},\ \bibinfo
  {pages} {205132} (\bibinfo {year} {2016})}\BibitemShut {NoStop}%
\bibitem [{\citenamefont {Jin}\ \emph {et~al.}(2019)\citenamefont {Jin},
  \citenamefont {Huang}, \citenamefont {Wang},\ and\ \citenamefont
  {Liu}}]{GlidePlane1}%
  \BibitemOpen
  \bibfield  {author} {\bibinfo {author} {\bibfnamefont {K.-H.}\ \bibnamefont
  {Jin}}, \bibinfo {author} {\bibfnamefont {H.}~\bibnamefont {Huang}}, \bibinfo
  {author} {\bibfnamefont {Z.}~\bibnamefont {Wang}}, \ and\ \bibinfo {author}
  {\bibfnamefont {F.}~\bibnamefont {Liu}},\ }\href {\doibase
  10.1039/C9NR00906J} {\bibfield  {journal} {\bibinfo  {journal} {Nanoscale}\
  }\textbf {\bibinfo {volume} {11}},\ \bibinfo {pages} {7256} (\bibinfo {year}
  {2019})}\BibitemShut {NoStop}%
\bibitem [{\citenamefont {Oh}\ \emph {et~al.}(2019)\citenamefont {Oh},
  \citenamefont {Min},\ and\ \citenamefont {Kim}}]{GlidePlane2}%
  \BibitemOpen
  \bibfield  {author} {\bibinfo {author} {\bibfnamefont {Y.-T.}\ \bibnamefont
  {Oh}}, \bibinfo {author} {\bibfnamefont {H.-G.}\ \bibnamefont {Min}}, \ and\
  \bibinfo {author} {\bibfnamefont {Y.}~\bibnamefont {Kim}},\ }\href {\doibase
  10.1103/PhysRevB.99.201110} {\bibfield  {journal} {\bibinfo  {journal} {Phys.
  Rev. B}\ }\textbf {\bibinfo {volume} {99}},\ \bibinfo {pages} {201110}
  (\bibinfo {year} {2019})}\BibitemShut {NoStop}%
\bibitem [{\citenamefont {Shao}\ \emph {et~al.}(2017)\citenamefont {Shao},
  \citenamefont {Chen}, \citenamefont {Gu}, \citenamefont {Guo}, \citenamefont
  {Lu}, \citenamefont {Sun}, \citenamefont {Sheng},\ and\ \citenamefont
  {Xing}}]{GlidePlane3}%
  \BibitemOpen
  \bibfield  {author} {\bibinfo {author} {\bibfnamefont {D.}~\bibnamefont
  {Shao}}, \bibinfo {author} {\bibfnamefont {T.}~\bibnamefont {Chen}}, \bibinfo
  {author} {\bibfnamefont {Q.}~\bibnamefont {Gu}}, \bibinfo {author}
  {\bibfnamefont {Z.}~\bibnamefont {Guo}}, \bibinfo {author} {\bibfnamefont
  {P.}~\bibnamefont {Lu}}, \bibinfo {author} {\bibfnamefont {J.}~\bibnamefont
  {Sun}}, \bibinfo {author} {\bibfnamefont {L.}~\bibnamefont {Sheng}}, \ and\
  \bibinfo {author} {\bibfnamefont {D.}~\bibnamefont {Xing}},\ }\href {\doibase
  10.1038/s41598-018-19870-5} {\bibfield  {journal} {\bibinfo  {journal}
  {Scientific Reports}\ }\textbf {\bibinfo {volume} {8}} (\bibinfo {year}
  {2017}),\ 10.1038/s41598-018-19870-5}\BibitemShut {NoStop}%
\bibitem [{\citenamefont {Ozawa}\ \emph {et~al.}(2019)\citenamefont {Ozawa},
  \citenamefont {Price}, \citenamefont {Amo}, \citenamefont {Goldman},
  \citenamefont {Hafezi}, \citenamefont {Lu}, \citenamefont {Rechtsman},
  \citenamefont {Schuster}, \citenamefont {Simon}, \citenamefont {Zilberberg},\
  and\ \citenamefont {Carusotto}}]{TopologicalPhotonics}%
  \BibitemOpen
  \bibfield  {author} {\bibinfo {author} {\bibfnamefont {T.}~\bibnamefont
  {Ozawa}}, \bibinfo {author} {\bibfnamefont {H.~M.}\ \bibnamefont {Price}},
  \bibinfo {author} {\bibfnamefont {A.}~\bibnamefont {Amo}}, \bibinfo {author}
  {\bibfnamefont {N.}~\bibnamefont {Goldman}}, \bibinfo {author} {\bibfnamefont
  {M.}~\bibnamefont {Hafezi}}, \bibinfo {author} {\bibfnamefont
  {L.}~\bibnamefont {Lu}}, \bibinfo {author} {\bibfnamefont {M.~C.}\
  \bibnamefont {Rechtsman}}, \bibinfo {author} {\bibfnamefont {D.}~\bibnamefont
  {Schuster}}, \bibinfo {author} {\bibfnamefont {J.}~\bibnamefont {Simon}},
  \bibinfo {author} {\bibfnamefont {O.}~\bibnamefont {Zilberberg}}, \ and\
  \bibinfo {author} {\bibfnamefont {I.}~\bibnamefont {Carusotto}},\ }\href
  {\doibase 10.1103/RevModPhys.91.015006} {\bibfield  {journal} {\bibinfo
  {journal} {Rev. Mod. Phys.}\ }\textbf {\bibinfo {volume} {91}},\ \bibinfo
  {pages} {015006} (\bibinfo {year} {2019})}\BibitemShut {NoStop}%
\bibitem [{\citenamefont {Gao}\ \emph {et~al.}(2018)\citenamefont {Gao},
  \citenamefont {Yang}, \citenamefont {Tremain}, \citenamefont {Liu},
  \citenamefont {Guo}, \citenamefont {Xia}, \citenamefont {Hibbins},\ and\
  \citenamefont {Zhang}}]{TopologicalPhotonics2}%
  \BibitemOpen
  \bibfield  {author} {\bibinfo {author} {\bibfnamefont {W.}~\bibnamefont
  {Gao}}, \bibinfo {author} {\bibfnamefont {B.}~\bibnamefont {Yang}}, \bibinfo
  {author} {\bibfnamefont {B.}~\bibnamefont {Tremain}}, \bibinfo {author}
  {\bibfnamefont {H.}~\bibnamefont {Liu}}, \bibinfo {author} {\bibfnamefont
  {Q.}~\bibnamefont {Guo}}, \bibinfo {author} {\bibfnamefont {L.}~\bibnamefont
  {Xia}}, \bibinfo {author} {\bibfnamefont {A.}~\bibnamefont {Hibbins}}, \ and\
  \bibinfo {author} {\bibfnamefont {S.}~\bibnamefont {Zhang}},\ }\href
  {\doibase 10.1038/s41467-018-03407-5} {\bibfield  {journal} {\bibinfo
  {journal} {Nature Communications}\ }\textbf {\bibinfo {volume} {9}} (\bibinfo
  {year} {2018}),\ 10.1038/s41467-018-03407-5}\BibitemShut {NoStop}%
\bibitem [{\citenamefont {Klembt}\ \emph {et~al.}(2018)\citenamefont {Klembt},
  \citenamefont {Harder}, \citenamefont {Egorov}, \citenamefont {Winkler},
  \citenamefont {Ge}, \citenamefont {Bandres}, \citenamefont {Emmerling},
  \citenamefont {Worschech}, \citenamefont {Liew}, \citenamefont {Segev},\ and\
  \citenamefont {et~al.}}]{TopologicalPolariton}%
  \BibitemOpen
  \bibfield  {author} {\bibinfo {author} {\bibfnamefont {S.}~\bibnamefont
  {Klembt}}, \bibinfo {author} {\bibfnamefont {T.~H.}\ \bibnamefont {Harder}},
  \bibinfo {author} {\bibfnamefont {O.~A.}\ \bibnamefont {Egorov}}, \bibinfo
  {author} {\bibfnamefont {K.}~\bibnamefont {Winkler}}, \bibinfo {author}
  {\bibfnamefont {R.}~\bibnamefont {Ge}}, \bibinfo {author} {\bibfnamefont
  {M.~A.}\ \bibnamefont {Bandres}}, \bibinfo {author} {\bibfnamefont
  {M.}~\bibnamefont {Emmerling}}, \bibinfo {author} {\bibfnamefont
  {L.}~\bibnamefont {Worschech}}, \bibinfo {author} {\bibfnamefont {T.~C.~H.}\
  \bibnamefont {Liew}}, \bibinfo {author} {\bibfnamefont {M.}~\bibnamefont
  {Segev}}, \ and\ \bibinfo {author} {\bibnamefont {et~al.}},\ }\href {\doibase
  10.1038/s41586-018-0601-5} {\bibfield  {journal} {\bibinfo  {journal}
  {Nature}\ }\textbf {\bibinfo {volume} {562}},\ \bibinfo {pages} {552–556}
  (\bibinfo {year} {2018})}\BibitemShut {NoStop}%
\bibitem [{\citenamefont {Karzig}\ \emph {et~al.}(2015)\citenamefont {Karzig},
  \citenamefont {Bardyn}, \citenamefont {Lindner},\ and\ \citenamefont
  {Refael}}]{TopologicalPolariton2}%
  \BibitemOpen
  \bibfield  {author} {\bibinfo {author} {\bibfnamefont {T.}~\bibnamefont
  {Karzig}}, \bibinfo {author} {\bibfnamefont {C.-E.}\ \bibnamefont {Bardyn}},
  \bibinfo {author} {\bibfnamefont {N.~H.}\ \bibnamefont {Lindner}}, \ and\
  \bibinfo {author} {\bibfnamefont {G.}~\bibnamefont {Refael}},\ }\href
  {\doibase 10.1103/PhysRevX.5.031001} {\bibfield  {journal} {\bibinfo
  {journal} {Phys. Rev. X}\ }\textbf {\bibinfo {volume} {5}},\ \bibinfo {pages}
  {031001} (\bibinfo {year} {2015})}\BibitemShut {NoStop}%
\bibitem [{\citenamefont {Owerre}(2016)}]{TopologicalMagnon}%
  \BibitemOpen
  \bibfield  {author} {\bibinfo {author} {\bibfnamefont {S.~A.}\ \bibnamefont
  {Owerre}},\ }\href {\doibase 10.1088/0953-8984/28/38/386001} {\bibfield
  {journal} {\bibinfo  {journal} {Journal of Physics: Condensed Matter}\
  }\textbf {\bibinfo {volume} {28}},\ \bibinfo {pages} {386001} (\bibinfo
  {year} {2016})}\BibitemShut {NoStop}%
\bibitem [{\citenamefont {Bhowmick}\ and\ \citenamefont
  {Sengupta}(2020)}]{TopologicalMagnon2}%
  \BibitemOpen
  \bibfield  {author} {\bibinfo {author} {\bibfnamefont {D.}~\bibnamefont
  {Bhowmick}}\ and\ \bibinfo {author} {\bibfnamefont {P.}~\bibnamefont
  {Sengupta}},\ }\href@noop {} {\enquote {\bibinfo {title} {The topological
  magnon bands in the flux state in sashtry-sutherland lattice},}\ } (\bibinfo
  {year} {2020}),\ \Eprint {http://arxiv.org/abs/2001.07008} {arXiv:2001.07008
  [cond-mat.str-el]} \BibitemShut {NoStop}%
\bibitem [{\citenamefont {Malz}\ \emph {et~al.}(2019)\citenamefont {Malz},
  \citenamefont {Knolle},\ and\ \citenamefont
  {Nunnenkamp}}]{TopologicalMagnon3}%
  \BibitemOpen
  \bibfield  {author} {\bibinfo {author} {\bibfnamefont {D.}~\bibnamefont
  {Malz}}, \bibinfo {author} {\bibfnamefont {J.}~\bibnamefont {Knolle}}, \ and\
  \bibinfo {author} {\bibfnamefont {A.}~\bibnamefont {Nunnenkamp}},\ }\href
  {\doibase 10.1038/s41467-019-11914-2} {\bibfield  {journal} {\bibinfo
  {journal} {Nature Communications}\ }\textbf {\bibinfo {volume} {10}}
  (\bibinfo {year} {2019}),\ 10.1038/s41467-019-11914-2}\BibitemShut {NoStop}%
\bibitem [{\citenamefont {Malki}\ and\ \citenamefont
  {Uhrig}(2019)}]{TopologicalMagnon4}%
  \BibitemOpen
  \bibfield  {author} {\bibinfo {author} {\bibfnamefont {M.}~\bibnamefont
  {Malki}}\ and\ \bibinfo {author} {\bibfnamefont {G.~S.}\ \bibnamefont
  {Uhrig}},\ }\href {\doibase 10.1103/PhysRevB.99.174412} {\bibfield  {journal}
  {\bibinfo  {journal} {Phys. Rev. B}\ }\textbf {\bibinfo {volume} {99}},\
  \bibinfo {pages} {174412} (\bibinfo {year} {2019})}\BibitemShut {NoStop}%
\bibitem [{\citenamefont {Kawano}\ and\ \citenamefont
  {Hotta}(2019)}]{TopologicalMagnon5}%
  \BibitemOpen
  \bibfield  {author} {\bibinfo {author} {\bibfnamefont {M.}~\bibnamefont
  {Kawano}}\ and\ \bibinfo {author} {\bibfnamefont {C.}~\bibnamefont {Hotta}},\
  }\href {\doibase 10.1103/PhysRevB.99.054422} {\bibfield  {journal} {\bibinfo
  {journal} {Phys. Rev. B}\ }\textbf {\bibinfo {volume} {99}},\ \bibinfo
  {pages} {054422} (\bibinfo {year} {2019})}\BibitemShut {NoStop}%
\bibitem [{\citenamefont {Joshi}(2018)}]{TopologicalMagnon6}%
  \BibitemOpen
  \bibfield  {author} {\bibinfo {author} {\bibfnamefont {D.}~\bibnamefont
  {Joshi}},\ }\href {\doibase 10.1103/PhysRevB.98.060405} {\bibfield  {journal}
  {\bibinfo  {journal} {Physical Review B}\ }\textbf {\bibinfo {volume} {98}}
  (\bibinfo {year} {2018}),\ 10.1103/PhysRevB.98.060405}\BibitemShut {NoStop}%
\bibitem [{\citenamefont {Joshi}\ and\ \citenamefont
  {Schnyder}(2019)}]{TopologicalMagnon7}%
  \BibitemOpen
  \bibfield  {author} {\bibinfo {author} {\bibfnamefont {D.}~\bibnamefont
  {Joshi}}\ and\ \bibinfo {author} {\bibfnamefont {A.}~\bibnamefont
  {Schnyder}},\ }\href {\doibase 10.1103/PhysRevB.100.020407} {\bibfield
  {journal} {\bibinfo  {journal} {Physical Review B}\ }\textbf {\bibinfo
  {volume} {100}} (\bibinfo {year} {2019}),\
  10.1103/PhysRevB.100.020407}\BibitemShut {NoStop}%
\bibitem [{\citenamefont {Lee}\ \emph {et~al.}(2018)\citenamefont {Lee},
  \citenamefont {Chung}, \citenamefont {Park},\ and\ \citenamefont
  {Park}}]{new1}%
  \BibitemOpen
  \bibfield  {author} {\bibinfo {author} {\bibfnamefont {K.~H.}\ \bibnamefont
  {Lee}}, \bibinfo {author} {\bibfnamefont {S.~B.}\ \bibnamefont {Chung}},
  \bibinfo {author} {\bibfnamefont {K.}~\bibnamefont {Park}}, \ and\ \bibinfo
  {author} {\bibfnamefont {J.-G.}\ \bibnamefont {Park}},\ }\href {\doibase
  10.1103/PhysRevB.97.180401} {\bibfield  {journal} {\bibinfo  {journal} {Phys.
  Rev. B}\ }\textbf {\bibinfo {volume} {97}},\ \bibinfo {pages} {180401}
  (\bibinfo {year} {2018})}\BibitemShut {NoStop}%
\bibitem [{\citenamefont {Kim}\ \emph {et~al.}(2016)\citenamefont {Kim},
  \citenamefont {Ochoa}, \citenamefont {Zarzuela},\ and\ \citenamefont
  {Tserkovnyak}}]{new2}%
  \BibitemOpen
  \bibfield  {author} {\bibinfo {author} {\bibfnamefont {S.~K.}\ \bibnamefont
  {Kim}}, \bibinfo {author} {\bibfnamefont {H.}~\bibnamefont {Ochoa}}, \bibinfo
  {author} {\bibfnamefont {R.}~\bibnamefont {Zarzuela}}, \ and\ \bibinfo
  {author} {\bibfnamefont {Y.}~\bibnamefont {Tserkovnyak}},\ }\href {\doibase
  10.1103/PhysRevLett.117.227201} {\bibfield  {journal} {\bibinfo  {journal}
  {Phys. Rev. Lett.}\ }\textbf {\bibinfo {volume} {117}},\ \bibinfo {pages}
  {227201} (\bibinfo {year} {2016})}\BibitemShut {NoStop}%
\bibitem [{\citenamefont {Bhowmick}\ and\ \citenamefont
  {Sengupta}(2019)}]{new3}%
  \BibitemOpen
  \bibfield  {author} {\bibinfo {author} {\bibfnamefont {D.}~\bibnamefont
  {Bhowmick}}\ and\ \bibinfo {author} {\bibfnamefont {P.}~\bibnamefont
  {Sengupta}},\ }\href@noop {} {\enquote {\bibinfo {title} {Anti-chiral edge
  states in heisenberg ferromagnet on a honeycomb lattice},}\ } (\bibinfo
  {year} {2019}),\ \Eprint {http://arxiv.org/abs/1908.04580} {arXiv:1908.04580
  [cond-mat.str-el]} \BibitemShut {NoStop}%
\bibitem [{\citenamefont {Chisnell}\ \emph {et~al.}(2015)\citenamefont
  {Chisnell}, \citenamefont {Helton}, \citenamefont {Freedman}, \citenamefont
  {Singh}, \citenamefont {Bewley}, \citenamefont {Nocera},\ and\ \citenamefont
  {Lee}}]{new4}%
  \BibitemOpen
  \bibfield  {author} {\bibinfo {author} {\bibfnamefont {R.}~\bibnamefont
  {Chisnell}}, \bibinfo {author} {\bibfnamefont {J.~S.}\ \bibnamefont
  {Helton}}, \bibinfo {author} {\bibfnamefont {D.~E.}\ \bibnamefont
  {Freedman}}, \bibinfo {author} {\bibfnamefont {D.~K.}\ \bibnamefont {Singh}},
  \bibinfo {author} {\bibfnamefont {R.~I.}\ \bibnamefont {Bewley}}, \bibinfo
  {author} {\bibfnamefont {D.~G.}\ \bibnamefont {Nocera}}, \ and\ \bibinfo
  {author} {\bibfnamefont {Y.~S.}\ \bibnamefont {Lee}},\ }\href {\doibase
  10.1103/PhysRevLett.115.147201} {\bibfield  {journal} {\bibinfo  {journal}
  {Phys. Rev. Lett.}\ }\textbf {\bibinfo {volume} {115}},\ \bibinfo {pages}
  {147201} (\bibinfo {year} {2015})}\BibitemShut {NoStop}%
\bibitem [{\citenamefont {Seshadri}\ and\ \citenamefont {Sen}(2018)}]{new5}%
  \BibitemOpen
  \bibfield  {author} {\bibinfo {author} {\bibfnamefont {R.}~\bibnamefont
  {Seshadri}}\ and\ \bibinfo {author} {\bibfnamefont {D.}~\bibnamefont {Sen}},\
  }\href {\doibase 10.1103/PhysRevB.97.134411} {\bibfield  {journal} {\bibinfo
  {journal} {Phys. Rev. B}\ }\textbf {\bibinfo {volume} {97}},\ \bibinfo
  {pages} {134411} (\bibinfo {year} {2018})}\BibitemShut {NoStop}%
\bibitem [{\citenamefont {Li}\ \emph {et~al.}(2016)\citenamefont {Li},
  \citenamefont {Li}, \citenamefont {Kim}, \citenamefont {Balents},
  \citenamefont {Yu},\ and\ \citenamefont {Chen}}]{WeylMagnon1}%
  \BibitemOpen
  \bibfield  {author} {\bibinfo {author} {\bibfnamefont {F.-Y.}\ \bibnamefont
  {Li}}, \bibinfo {author} {\bibfnamefont {Y.-D.}\ \bibnamefont {Li}}, \bibinfo
  {author} {\bibfnamefont {Y.~B.}\ \bibnamefont {Kim}}, \bibinfo {author}
  {\bibfnamefont {L.}~\bibnamefont {Balents}}, \bibinfo {author} {\bibfnamefont
  {Y.}~\bibnamefont {Yu}}, \ and\ \bibinfo {author} {\bibfnamefont
  {G.}~\bibnamefont {Chen}},\ }\href {\doibase 10.1038/ncomms12691} {\bibfield
  {journal} {\bibinfo  {journal} {Nature Communications}\ }\textbf {\bibinfo
  {volume} {7}} (\bibinfo {year} {2016}),\ 10.1038/ncomms12691}\BibitemShut
  {NoStop}%
\bibitem [{\citenamefont {Li}\ and\ \citenamefont {Hu}(2017)}]{WeylMagnon2}%
  \BibitemOpen
  \bibfield  {author} {\bibinfo {author} {\bibfnamefont {K.-K.}\ \bibnamefont
  {Li}}\ and\ \bibinfo {author} {\bibfnamefont {J.-P.}\ \bibnamefont {Hu}},\
  }\href {\doibase 10.1088/0256-307x/34/7/077501} {\bibfield  {journal}
  {\bibinfo  {journal} {Chinese Physics Letters}\ }\textbf {\bibinfo {volume}
  {34}},\ \bibinfo {pages} {077501} (\bibinfo {year} {2017})}\BibitemShut
  {NoStop}%
\bibitem [{\citenamefont {Liu}\ and\ \citenamefont {Shi}(2019)}]{WeylMagnon3}%
  \BibitemOpen
  \bibfield  {author} {\bibinfo {author} {\bibfnamefont {T.}~\bibnamefont
  {Liu}}\ and\ \bibinfo {author} {\bibfnamefont {Z.}~\bibnamefont {Shi}},\
  }\href {\doibase 10.1103/physrevb.99.214413} {\bibfield  {journal} {\bibinfo
  {journal} {Physical Review B}\ }\textbf {\bibinfo {volume} {99}} (\bibinfo
  {year} {2019}),\ 10.1103/physrevb.99.214413}\BibitemShut {NoStop}%
\bibitem [{\citenamefont {Owerre}(2018)}]{WeylMagnon4}%
  \BibitemOpen
  \bibfield  {author} {\bibinfo {author} {\bibfnamefont {S.~A.}\ \bibnamefont
  {Owerre}},\ }\href {\doibase 10.1103/PhysRevB.97.094412} {\bibfield
  {journal} {\bibinfo  {journal} {Phys. Rev. B}\ }\textbf {\bibinfo {volume}
  {97}},\ \bibinfo {pages} {094412} (\bibinfo {year} {2018})}\BibitemShut
  {NoStop}%
\bibitem [{\citenamefont {Jian}\ and\ \citenamefont {Nie}(2018)}]{WeylMagnon5}%
  \BibitemOpen
  \bibfield  {author} {\bibinfo {author} {\bibfnamefont {S.-K.}\ \bibnamefont
  {Jian}}\ and\ \bibinfo {author} {\bibfnamefont {W.}~\bibnamefont {Nie}},\
  }\href {\doibase 10.1103/PhysRevB.97.115162} {\bibfield  {journal} {\bibinfo
  {journal} {Phys. Rev. B}\ }\textbf {\bibinfo {volume} {97}},\ \bibinfo
  {pages} {115162} (\bibinfo {year} {2018})}\BibitemShut {NoStop}%
\bibitem [{\citenamefont {Owerre}(2019{\natexlab{a}})}]{WeylMagnon6}%
  \BibitemOpen
  \bibfield  {author} {\bibinfo {author} {\bibfnamefont {S.}~\bibnamefont
  {Owerre}},\ }\href {\doibase https://doi.org/10.1016/j.aop.2019.04.003}
  {\bibfield  {journal} {\bibinfo  {journal} {Annals of Physics}\ }\textbf
  {\bibinfo {volume} {406}},\ \bibinfo {pages} {14 } (\bibinfo {year}
  {2019}{\natexlab{a}})}\BibitemShut {NoStop}%
\bibitem [{\citenamefont {Shen}\ and\ \citenamefont
  {Kim}(2020)}]{MagnonPolaron1}%
  \BibitemOpen
  \bibfield  {author} {\bibinfo {author} {\bibfnamefont {P.}~\bibnamefont
  {Shen}}\ and\ \bibinfo {author} {\bibfnamefont {S.~K.}\ \bibnamefont {Kim}},\
  }\href {\doibase 10.1103/PhysRevB.101.125111} {\bibfield  {journal} {\bibinfo
   {journal} {Phys. Rev. B}\ }\textbf {\bibinfo {volume} {101}},\ \bibinfo
  {pages} {125111} (\bibinfo {year} {2020})}\BibitemShut {NoStop}%
\bibitem [{\citenamefont {Park}\ and\ \citenamefont
  {Yang}(2019)}]{MagnonPolaron2}%
  \BibitemOpen
  \bibfield  {author} {\bibinfo {author} {\bibfnamefont {S.}~\bibnamefont
  {Park}}\ and\ \bibinfo {author} {\bibfnamefont {B.-J.}\ \bibnamefont
  {Yang}},\ }\href {\doibase 10.1103/PhysRevB.99.174435} {\bibfield  {journal}
  {\bibinfo  {journal} {Phys. Rev. B}\ }\textbf {\bibinfo {volume} {99}},\
  \bibinfo {pages} {174435} (\bibinfo {year} {2019})}\BibitemShut {NoStop}%
\bibitem [{\citenamefont {Zhang}\ \emph {et~al.}(2019)\citenamefont {Zhang},
  \citenamefont {Go}, \citenamefont {Lee},\ and\ \citenamefont
  {Kim}}]{MagnonPolaron3}%
  \BibitemOpen
  \bibfield  {author} {\bibinfo {author} {\bibfnamefont {S.}~\bibnamefont
  {Zhang}}, \bibinfo {author} {\bibfnamefont {G.}~\bibnamefont {Go}}, \bibinfo
  {author} {\bibfnamefont {K.-J.}\ \bibnamefont {Lee}}, \ and\ \bibinfo
  {author} {\bibfnamefont {S.~K.}\ \bibnamefont {Kim}},\ }\href@noop {}
  {\enquote {\bibinfo {title} {Su(3) topology of magnon-phonon hybridization in
  2d antiferromagnets},}\ } (\bibinfo {year} {2019}),\ \Eprint
  {http://arxiv.org/abs/1909.08031} {arXiv:1909.08031 [cond-mat.mes-hall]}
  \BibitemShut {NoStop}%
\bibitem [{\citenamefont {Go}\ \emph {et~al.}(2019)\citenamefont {Go},
  \citenamefont {Kim},\ and\ \citenamefont {Lee}}]{MagnonPolaron4}%
  \BibitemOpen
  \bibfield  {author} {\bibinfo {author} {\bibfnamefont {G.}~\bibnamefont
  {Go}}, \bibinfo {author} {\bibfnamefont {S.~K.}\ \bibnamefont {Kim}}, \ and\
  \bibinfo {author} {\bibfnamefont {K.-J.}\ \bibnamefont {Lee}},\ }\href
  {\doibase 10.1103/PhysRevLett.123.237207} {\bibfield  {journal} {\bibinfo
  {journal} {Phys. Rev. Lett.}\ }\textbf {\bibinfo {volume} {123}},\ \bibinfo
  {pages} {237207} (\bibinfo {year} {2019})}\BibitemShut {NoStop}%
\bibitem [{\citenamefont {Mook}\ \emph {et~al.}(2017)\citenamefont {Mook},
  \citenamefont {Henk},\ and\ \citenamefont {Mertig}}]{PyrochloreFerromagnet}%
  \BibitemOpen
  \bibfield  {author} {\bibinfo {author} {\bibfnamefont {A.}~\bibnamefont
  {Mook}}, \bibinfo {author} {\bibfnamefont {J.}~\bibnamefont {Henk}}, \ and\
  \bibinfo {author} {\bibfnamefont {I.}~\bibnamefont {Mertig}},\ }\href
  {\doibase 10.1103/PhysRevB.95.014418} {\bibfield  {journal} {\bibinfo
  {journal} {Phys. Rev. B}\ }\textbf {\bibinfo {volume} {95}},\ \bibinfo
  {pages} {014418} (\bibinfo {year} {2017})}\BibitemShut {NoStop}%
\bibitem [{\citenamefont {Li}\ \emph {et~al.}(2017)\citenamefont {Li},
  \citenamefont {Li}, \citenamefont {Hu}, \citenamefont {Li},\ and\
  \citenamefont {Fang}}]{SpinWeb}%
  \BibitemOpen
  \bibfield  {author} {\bibinfo {author} {\bibfnamefont {K.}~\bibnamefont
  {Li}}, \bibinfo {author} {\bibfnamefont {C.}~\bibnamefont {Li}}, \bibinfo
  {author} {\bibfnamefont {J.}~\bibnamefont {Hu}}, \bibinfo {author}
  {\bibfnamefont {Y.}~\bibnamefont {Li}}, \ and\ \bibinfo {author}
  {\bibfnamefont {C.}~\bibnamefont {Fang}},\ }\href {\doibase
  10.1103/PhysRevLett.119.247202} {\bibfield  {journal} {\bibinfo  {journal}
  {Phys. Rev. Lett.}\ }\textbf {\bibinfo {volume} {119}},\ \bibinfo {pages}
  {247202} (\bibinfo {year} {2017})}\BibitemShut {NoStop}%
\bibitem [{\citenamefont
  {Owerre}(2019{\natexlab{b}})}]{LayeredAntiFerromagnet}%
  \BibitemOpen
  \bibfield  {author} {\bibinfo {author} {\bibfnamefont {S.~A.}\ \bibnamefont
  {Owerre}},\ }\href {\doibase 10.1209/0295-5075/125/36002} {\bibfield
  {journal} {\bibinfo  {journal} {EPL (Europhysics Letters)}\ }\textbf
  {\bibinfo {volume} {125}},\ \bibinfo {pages} {36002} (\bibinfo {year}
  {2019}{\natexlab{b}})}\BibitemShut {NoStop}%
\bibitem [{\citenamefont {Petersen}\ and\ \citenamefont
  {Hedegård}(2000)}]{2DSemimetal}%
  \BibitemOpen
  \bibfield  {author} {\bibinfo {author} {\bibfnamefont {L.}~\bibnamefont
  {Petersen}}\ and\ \bibinfo {author} {\bibfnamefont {P.}~\bibnamefont
  {Hedegård}},\ }\href {\doibase
  https://doi.org/10.1016/S0039-6028(00)00441-6} {\bibfield  {journal}
  {\bibinfo  {journal} {Surface Science}\ }\textbf {\bibinfo {volume} {459}},\
  \bibinfo {pages} {49 } (\bibinfo {year} {2000})}\BibitemShut {NoStop}%
\bibitem [{\citenamefont {Lu}\ \emph {et~al.}(2017)\citenamefont {Lu},
  \citenamefont {Luo}, \citenamefont {Li}, \citenamefont {Yang}, \citenamefont
  {Cao}, \citenamefont {Gong},\ and\ \citenamefont {Xiang}}]{2DSemimetal2}%
  \BibitemOpen
  \bibfield  {author} {\bibinfo {author} {\bibfnamefont {J.-L.}\ \bibnamefont
  {Lu}}, \bibinfo {author} {\bibfnamefont {W.}~\bibnamefont {Luo}}, \bibinfo
  {author} {\bibfnamefont {X.-Y.}\ \bibnamefont {Li}}, \bibinfo {author}
  {\bibfnamefont {S.-Q.}\ \bibnamefont {Yang}}, \bibinfo {author}
  {\bibfnamefont {J.-X.}\ \bibnamefont {Cao}}, \bibinfo {author} {\bibfnamefont
  {X.-G.}\ \bibnamefont {Gong}}, \ and\ \bibinfo {author} {\bibfnamefont
  {H.-J.}\ \bibnamefont {Xiang}},\ }\href {\doibase
  10.1088/0256-307x/34/5/057302} {\bibfield  {journal} {\bibinfo  {journal}
  {Chinese Physics Letters}\ }\textbf {\bibinfo {volume} {34}},\ \bibinfo
  {pages} {057302} (\bibinfo {year} {2017})}\BibitemShut {NoStop}%
\bibitem [{\citenamefont {Jin}\ \emph {et~al.}(2017)\citenamefont {Jin},
  \citenamefont {Wang}, \citenamefont {Zhao}, \citenamefont {Du}, \citenamefont
  {Zheng}, \citenamefont {Gan}, \citenamefont {Liu}, \citenamefont {Xu},\ and\
  \citenamefont {Tong}}]{2DSemimetal3}%
  \BibitemOpen
  \bibfield  {author} {\bibinfo {author} {\bibfnamefont {Y.-J.}\ \bibnamefont
  {Jin}}, \bibinfo {author} {\bibfnamefont {R.}~\bibnamefont {Wang}}, \bibinfo
  {author} {\bibfnamefont {J.-Z.}\ \bibnamefont {Zhao}}, \bibinfo {author}
  {\bibfnamefont {Y.-P.}\ \bibnamefont {Du}}, \bibinfo {author} {\bibfnamefont
  {C.-D.}\ \bibnamefont {Zheng}}, \bibinfo {author} {\bibfnamefont {L.-Y.}\
  \bibnamefont {Gan}}, \bibinfo {author} {\bibfnamefont {J.-F.}\ \bibnamefont
  {Liu}}, \bibinfo {author} {\bibfnamefont {H.}~\bibnamefont {Xu}}, \ and\
  \bibinfo {author} {\bibfnamefont {S.~Y.}\ \bibnamefont {Tong}},\ }\href
  {\doibase 10.1039/C7NR03520A} {\bibfield  {journal} {\bibinfo  {journal}
  {Nanoscale}\ }\textbf {\bibinfo {volume} {9}},\ \bibinfo {pages} {13112}
  (\bibinfo {year} {2017})}\BibitemShut {NoStop}%
\bibitem [{\citenamefont {Feng}\ \emph {et~al.}(2017)\citenamefont {Feng},
  \citenamefont {Botao}, \citenamefont {Kasamatsu}, \citenamefont {Ito},
  \citenamefont {Cheng}, \citenamefont {Liu}, \citenamefont {Feng},
  \citenamefont {Wu}, \citenamefont {Mahatha}, \citenamefont {Sheverdyaeva},
  \citenamefont {Moras}, \citenamefont {Arita}, \citenamefont {Sugino},
  \citenamefont {Chiang}, \citenamefont {Shimada}, \citenamefont {Miyamoto},
  \citenamefont {Okuda}, \citenamefont {Wu}, \citenamefont {Chen},\ and\
  \citenamefont {Matsuda}}]{2DSemimetal4}%
  \BibitemOpen
  \bibfield  {author} {\bibinfo {author} {\bibfnamefont {B.}~\bibnamefont
  {Feng}}, \bibinfo {author} {\bibfnamefont {F.}~\bibnamefont {Botao}},
  \bibinfo {author} {\bibfnamefont {S.}~\bibnamefont {Kasamatsu}}, \bibinfo
  {author} {\bibfnamefont {S.}~\bibnamefont {Ito}}, \bibinfo {author}
  {\bibfnamefont {P.}~\bibnamefont {Cheng}}, \bibinfo {author} {\bibfnamefont
  {C.}~\bibnamefont {Liu}}, \bibinfo {author} {\bibfnamefont {Y.}~\bibnamefont
  {Feng}}, \bibinfo {author} {\bibfnamefont {S.}~\bibnamefont {Wu}}, \bibinfo
  {author} {\bibfnamefont {S.}~\bibnamefont {Mahatha}}, \bibinfo {author}
  {\bibfnamefont {P.}~\bibnamefont {Sheverdyaeva}}, \bibinfo {author}
  {\bibfnamefont {P.}~\bibnamefont {Moras}}, \bibinfo {author} {\bibfnamefont
  {M.}~\bibnamefont {Arita}}, \bibinfo {author} {\bibfnamefont
  {O.}~\bibnamefont {Sugino}}, \bibinfo {author} {\bibfnamefont {T.-C.}\
  \bibnamefont {Chiang}}, \bibinfo {author} {\bibfnamefont {K.}~\bibnamefont
  {Shimada}}, \bibinfo {author} {\bibfnamefont {K.}~\bibnamefont {Miyamoto}},
  \bibinfo {author} {\bibfnamefont {T.}~\bibnamefont {Okuda}}, \bibinfo
  {author} {\bibfnamefont {K.}~\bibnamefont {Wu}}, \bibinfo {author}
  {\bibfnamefont {L.}~\bibnamefont {Chen}}, \ and\ \bibinfo {author}
  {\bibfnamefont {I.}~\bibnamefont {Matsuda}},\ }\href {\doibase
  10.1038/s41467-017-01108-z} {\bibfield  {journal} {\bibinfo  {journal}
  {Nature Communications}\ }\textbf {\bibinfo {volume} {8}},\ \bibinfo {pages}
  {1007} (\bibinfo {year} {2017})}\BibitemShut {NoStop}%
\bibitem [{\citenamefont {Owerre}(2017)}]{Quasi2D}%
  \BibitemOpen
  \bibfield  {author} {\bibinfo {author} {\bibfnamefont {S.~A.}\ \bibnamefont
  {Owerre}},\ }\href {\doibase 10.1038/s41598-017-07276-8} {\bibfield
  {journal} {\bibinfo  {journal} {Scientific Reports}\ }\textbf {\bibinfo
  {volume} {7}} (\bibinfo {year} {2017}),\
  10.1038/s41598-017-07276-8}\BibitemShut {NoStop}%
\bibitem [{\citenamefont {Shastry}\ and\ \citenamefont
  {Sutherland}(1981)}]{Shastry}%
  \BibitemOpen
  \bibfield  {author} {\bibinfo {author} {\bibfnamefont {B.~S.}\ \bibnamefont
  {Shastry}}\ and\ \bibinfo {author} {\bibfnamefont {B.}~\bibnamefont
  {Sutherland}},\ }\href {\doibase
  https://doi.org/10.1016/0378-4363(81)90838-X} {\bibfield  {journal} {\bibinfo
   {journal} {Physica B+C}\ }\textbf {\bibinfo {volume} {108}},\ \bibinfo
  {pages} {1069 } (\bibinfo {year} {1981})}\BibitemShut {NoStop}%
\bibitem [{\citenamefont {Smith}\ and\ \citenamefont
  {Keszler}(1991)}]{structure1}%
  \BibitemOpen
  \bibfield  {author} {\bibinfo {author} {\bibfnamefont {R.~W.}\ \bibnamefont
  {Smith}}\ and\ \bibinfo {author} {\bibfnamefont {D.~A.}\ \bibnamefont
  {Keszler}},\ }\href@noop {} {\bibfield  {journal} {\bibinfo  {journal}
  {Journal of Solid State Chemistry}\ }\textbf {\bibinfo {volume} {93}},\
  \bibinfo {pages} {430} (\bibinfo {year} {1991})}\BibitemShut {NoStop}%
\bibitem [{\citenamefont {Sparta}\ \emph {et~al.}(2001)\citenamefont {Sparta},
  \citenamefont {Redhammer}, \citenamefont {Roussel}, \citenamefont {Heger},
  \citenamefont {Roth}, \citenamefont {Lemmens}, \citenamefont {Ionescu},
  \citenamefont {Grove}, \citenamefont {G{\"u}ntherodt}, \citenamefont
  {H{\"u}ning} \emph {et~al.}}]{structure2}%
  \BibitemOpen
  \bibfield  {author} {\bibinfo {author} {\bibfnamefont {K.}~\bibnamefont
  {Sparta}}, \bibinfo {author} {\bibfnamefont {G.}~\bibnamefont {Redhammer}},
  \bibinfo {author} {\bibfnamefont {P.}~\bibnamefont {Roussel}}, \bibinfo
  {author} {\bibfnamefont {G.}~\bibnamefont {Heger}}, \bibinfo {author}
  {\bibfnamefont {G.}~\bibnamefont {Roth}}, \bibinfo {author} {\bibfnamefont
  {P.}~\bibnamefont {Lemmens}}, \bibinfo {author} {\bibfnamefont
  {A.}~\bibnamefont {Ionescu}}, \bibinfo {author} {\bibfnamefont
  {M.}~\bibnamefont {Grove}}, \bibinfo {author} {\bibfnamefont
  {G.}~\bibnamefont {G{\"u}ntherodt}}, \bibinfo {author} {\bibfnamefont
  {F.}~\bibnamefont {H{\"u}ning}},  \emph {et~al.},\ }\href@noop {} {\bibfield
  {journal} {\bibinfo  {journal} {The European Physical Journal B-Condensed
  Matter and Complex Systems}\ }\textbf {\bibinfo {volume} {19}},\ \bibinfo
  {pages} {507} (\bibinfo {year} {2001})}\BibitemShut {NoStop}%
\bibitem [{\citenamefont {Romh\'anyi}\ \emph {et~al.}(2011)\citenamefont
  {Romh\'anyi}, \citenamefont {Totsuka},\ and\ \citenamefont {Penc}}]{phase4}%
  \BibitemOpen
  \bibfield  {author} {\bibinfo {author} {\bibfnamefont {J.}~\bibnamefont
  {Romh\'anyi}}, \bibinfo {author} {\bibfnamefont {K.}~\bibnamefont {Totsuka}},
  \ and\ \bibinfo {author} {\bibfnamefont {K.}~\bibnamefont {Penc}},\ }\href
  {\doibase 10.1103/PhysRevB.83.024413} {\bibfield  {journal} {\bibinfo
  {journal} {Phys. Rev. B}\ }\textbf {\bibinfo {volume} {83}},\ \bibinfo
  {pages} {024413} (\bibinfo {year} {2011})}\BibitemShut {NoStop}%
\bibitem [{\citenamefont {Miyahara}\ and\ \citenamefont {Ueda}(1999)}]{phase1}%
  \BibitemOpen
  \bibfield  {author} {\bibinfo {author} {\bibfnamefont {S.}~\bibnamefont
  {Miyahara}}\ and\ \bibinfo {author} {\bibfnamefont {K.}~\bibnamefont
  {Ueda}},\ }\href {\doibase 10.1103/PhysRevLett.82.3701} {\bibfield  {journal}
  {\bibinfo  {journal} {Phys. Rev. Lett.}\ }\textbf {\bibinfo {volume} {82}},\
  \bibinfo {pages} {3701} (\bibinfo {year} {1999})}\BibitemShut {NoStop}%
\bibitem [{\citenamefont {Koga}\ and\ \citenamefont {Kawakami}(2000)}]{phase2}%
  \BibitemOpen
  \bibfield  {author} {\bibinfo {author} {\bibfnamefont {A.}~\bibnamefont
  {Koga}}\ and\ \bibinfo {author} {\bibfnamefont {N.}~\bibnamefont
  {Kawakami}},\ }\href {\doibase 10.1103/PhysRevLett.84.4461} {\bibfield
  {journal} {\bibinfo  {journal} {Phys. Rev. Lett.}\ }\textbf {\bibinfo
  {volume} {84}},\ \bibinfo {pages} {4461} (\bibinfo {year}
  {2000})}\BibitemShut {NoStop}%
\bibitem [{\citenamefont {L\"auchli}\ \emph {et~al.}(2002)\citenamefont
  {L\"auchli}, \citenamefont {Wessel},\ and\ \citenamefont {Sigrist}}]{phase3}%
  \BibitemOpen
  \bibfield  {author} {\bibinfo {author} {\bibfnamefont {A.}~\bibnamefont
  {L\"auchli}}, \bibinfo {author} {\bibfnamefont {S.}~\bibnamefont {Wessel}}, \
  and\ \bibinfo {author} {\bibfnamefont {M.}~\bibnamefont {Sigrist}},\ }\href
  {\doibase 10.1103/PhysRevB.66.014401} {\bibfield  {journal} {\bibinfo
  {journal} {Phys. Rev. B}\ }\textbf {\bibinfo {volume} {66}},\ \bibinfo
  {pages} {014401} (\bibinfo {year} {2002})}\BibitemShut {NoStop}%
\bibitem [{\citenamefont {{Romh{\'a}nyi}}\ \emph {et~al.}(2015)\citenamefont
  {{Romh{\'a}nyi}}, \citenamefont {{Penc}},\ and\ \citenamefont
  {{Ganesh}}}]{triplon}%
  \BibitemOpen
  \bibfield  {author} {\bibinfo {author} {\bibfnamefont {J.}~\bibnamefont
  {{Romh{\'a}nyi}}}, \bibinfo {author} {\bibfnamefont {K.}~\bibnamefont
  {{Penc}}}, \ and\ \bibinfo {author} {\bibfnamefont {R.}~\bibnamefont
  {{Ganesh}}},\ }\href {\doibase 10.1038/ncomms7805} {\bibfield  {journal}
  {\bibinfo  {journal} {Nature Communications}\ }\textbf {\bibinfo {volume}
  {6}},\ \bibinfo {eid} {6805} (\bibinfo {year} {2015})},\ \Eprint
  {http://arxiv.org/abs/1406.1163} {arXiv:1406.1163 [cond-mat.str-el]}
  \BibitemShut {NoStop}%
\bibitem [{\citenamefont {Malki}\ and\ \citenamefont
  {Schmidt}(2017)}]{triplon2}%
  \BibitemOpen
  \bibfield  {author} {\bibinfo {author} {\bibfnamefont {M.}~\bibnamefont
  {Malki}}\ and\ \bibinfo {author} {\bibfnamefont {K.~P.}\ \bibnamefont
  {Schmidt}},\ }\href {\doibase 10.1103/PhysRevB.95.195137} {\bibfield
  {journal} {\bibinfo  {journal} {Phys. Rev. B}\ }\textbf {\bibinfo {volume}
  {95}},\ \bibinfo {pages} {195137} (\bibinfo {year} {2017})}\BibitemShut
  {NoStop}%
\bibitem [{\citenamefont {{McClarty}}\ \emph {et~al.}(2017)\citenamefont
  {{McClarty}}, \citenamefont {{Kr{\"u}ger}}, \citenamefont {{Guidi}},
  \citenamefont {{Parker}}, \citenamefont {{Refson}}, \citenamefont {{Parker}},
  \citenamefont {{Prabhakaran}},\ and\ \citenamefont {{Coldea}}}]{triplon3}%
  \BibitemOpen
  \bibfield  {author} {\bibinfo {author} {\bibfnamefont {P.~A.}\ \bibnamefont
  {{McClarty}}}, \bibinfo {author} {\bibfnamefont {F.}~\bibnamefont
  {{Kr{\"u}ger}}}, \bibinfo {author} {\bibfnamefont {T.}~\bibnamefont
  {{Guidi}}}, \bibinfo {author} {\bibfnamefont {S.~F.}\ \bibnamefont
  {{Parker}}}, \bibinfo {author} {\bibfnamefont {K.}~\bibnamefont {{Refson}}},
  \bibinfo {author} {\bibfnamefont {A.~W.}\ \bibnamefont {{Parker}}}, \bibinfo
  {author} {\bibfnamefont {D.}~\bibnamefont {{Prabhakaran}}}, \ and\ \bibinfo
  {author} {\bibfnamefont {R.}~\bibnamefont {{Coldea}}},\ }\href {\doibase
  10.1038/nphys4117} {\bibfield  {journal} {\bibinfo  {journal} {Nature
  Physics}\ }\textbf {\bibinfo {volume} {13}},\ \bibinfo {pages} {736}
  (\bibinfo {year} {2017})},\ \Eprint {http://arxiv.org/abs/1609.01922}
  {arXiv:1609.01922 [cond-mat.str-el]} \BibitemShut {NoStop}%
\bibitem [{\citenamefont {Kageyama}\ \emph {et~al.}(1999)\citenamefont
  {Kageyama}, \citenamefont {Yoshimura}, \citenamefont {Stern}, \citenamefont
  {Mushnikov}, \citenamefont {Onizuka}, \citenamefont {Kato}, \citenamefont
  {Kosuge}, \citenamefont {Slichter}, \citenamefont {Goto},\ and\ \citenamefont
  {Ueda}}]{dimer1}%
  \BibitemOpen
  \bibfield  {author} {\bibinfo {author} {\bibfnamefont {H.}~\bibnamefont
  {Kageyama}}, \bibinfo {author} {\bibfnamefont {K.}~\bibnamefont {Yoshimura}},
  \bibinfo {author} {\bibfnamefont {R.}~\bibnamefont {Stern}}, \bibinfo
  {author} {\bibfnamefont {N.~V.}\ \bibnamefont {Mushnikov}}, \bibinfo {author}
  {\bibfnamefont {K.}~\bibnamefont {Onizuka}}, \bibinfo {author} {\bibfnamefont
  {M.}~\bibnamefont {Kato}}, \bibinfo {author} {\bibfnamefont {K.}~\bibnamefont
  {Kosuge}}, \bibinfo {author} {\bibfnamefont {C.~P.}\ \bibnamefont
  {Slichter}}, \bibinfo {author} {\bibfnamefont {T.}~\bibnamefont {Goto}}, \
  and\ \bibinfo {author} {\bibfnamefont {Y.}~\bibnamefont {Ueda}},\ }\href
  {\doibase 10.1103/PhysRevLett.82.3168} {\bibfield  {journal} {\bibinfo
  {journal} {Phys. Rev. Lett.}\ }\textbf {\bibinfo {volume} {82}},\ \bibinfo
  {pages} {3168} (\bibinfo {year} {1999})}\BibitemShut {NoStop}%
\bibitem [{\citenamefont {Knetter}\ \emph {et~al.}(2000)\citenamefont
  {Knetter}, \citenamefont {B\"uhler}, \citenamefont {M\"uller-Hartmann},\ and\
  \citenamefont {Uhrig}}]{dimer3}%
  \BibitemOpen
  \bibfield  {author} {\bibinfo {author} {\bibfnamefont {C.}~\bibnamefont
  {Knetter}}, \bibinfo {author} {\bibfnamefont {A.}~\bibnamefont {B\"uhler}},
  \bibinfo {author} {\bibfnamefont {E.}~\bibnamefont {M\"uller-Hartmann}}, \
  and\ \bibinfo {author} {\bibfnamefont {G.~S.}\ \bibnamefont {Uhrig}},\ }\href
  {\doibase 10.1103/PhysRevLett.85.3958} {\bibfield  {journal} {\bibinfo
  {journal} {Phys. Rev. Lett.}\ }\textbf {\bibinfo {volume} {85}},\ \bibinfo
  {pages} {3958} (\bibinfo {year} {2000})}\BibitemShut {NoStop}%
\bibitem [{\citenamefont {Miyahara}\ and\ \citenamefont {Ueda}(2000)}]{dimer4}%
  \BibitemOpen
  \bibfield  {author} {\bibinfo {author} {\bibfnamefont {S.}~\bibnamefont
  {Miyahara}}\ and\ \bibinfo {author} {\bibfnamefont {K.}~\bibnamefont
  {Ueda}},\ }\href {\doibase 10.1103/PhysRevB.61.3417} {\bibfield  {journal}
  {\bibinfo  {journal} {Phys. Rev. B}\ }\textbf {\bibinfo {volume} {61}},\
  \bibinfo {pages} {3417} (\bibinfo {year} {2000})}\BibitemShut {NoStop}%
\bibitem [{\citenamefont {Sachdev}\ and\ \citenamefont
  {Bhatt}(1990)}]{BondOperator}%
  \BibitemOpen
  \bibfield  {author} {\bibinfo {author} {\bibfnamefont {S.}~\bibnamefont
  {Sachdev}}\ and\ \bibinfo {author} {\bibfnamefont {R.~N.}\ \bibnamefont
  {Bhatt}},\ }\href {\doibase 10.1103/PhysRevB.41.9323} {\bibfield  {journal}
  {\bibinfo  {journal} {Phys. Rev. B}\ }\textbf {\bibinfo {volume} {41}},\
  \bibinfo {pages} {9323} (\bibinfo {year} {1990})}\BibitemShut {NoStop}%
\bibitem [{\citenamefont {Nojiri}\ \emph {et~al.}(2003)\citenamefont {Nojiri},
  \citenamefont {Kageyama}, \citenamefont {Ueda},\ and\ \citenamefont
  {Motokawa}}]{ESR}%
  \BibitemOpen
  \bibfield  {author} {\bibinfo {author} {\bibfnamefont {H.}~\bibnamefont
  {Nojiri}}, \bibinfo {author} {\bibfnamefont {H.}~\bibnamefont {Kageyama}},
  \bibinfo {author} {\bibfnamefont {Y.}~\bibnamefont {Ueda}}, \ and\ \bibinfo
  {author} {\bibfnamefont {M.}~\bibnamefont {Motokawa}},\ }\href@noop {}
  {\bibfield  {journal} {\bibinfo  {journal} {Journal of the Physical Society
  of Japan}\ }\textbf {\bibinfo {volume} {72}},\ \bibinfo {pages} {3243}
  (\bibinfo {year} {2003})}\BibitemShut {NoStop}%
\bibitem [{\citenamefont {T\"urker}\ and\ \citenamefont
  {Moroz}(2018)}]{symmetry1}%
  \BibitemOpen
  \bibfield  {author} {\bibinfo {author} {\bibfnamefont {O.~b.~u.}\
  \bibnamefont {T\"urker}}\ and\ \bibinfo {author} {\bibfnamefont
  {S.}~\bibnamefont {Moroz}},\ }\href {\doibase 10.1103/PhysRevB.97.075120}
  {\bibfield  {journal} {\bibinfo  {journal} {Phys. Rev. B}\ }\textbf {\bibinfo
  {volume} {97}},\ \bibinfo {pages} {075120} (\bibinfo {year}
  {2018})}\BibitemShut {NoStop}%
\bibitem [{\citenamefont {Fang}\ \emph {et~al.}(2016)\citenamefont {Fang},
  \citenamefont {Weng}, \citenamefont {Dai},\ and\ \citenamefont
  {Fang}}]{symmetry2}%
  \BibitemOpen
  \bibfield  {author} {\bibinfo {author} {\bibfnamefont {C.}~\bibnamefont
  {Fang}}, \bibinfo {author} {\bibfnamefont {H.}~\bibnamefont {Weng}}, \bibinfo
  {author} {\bibfnamefont {X.}~\bibnamefont {Dai}}, \ and\ \bibinfo {author}
  {\bibfnamefont {Z.}~\bibnamefont {Fang}},\ }\href {\doibase
  10.1088/1674-1056/25/11/117106} {\bibfield  {journal} {\bibinfo  {journal}
  {Chinese Physics B}\ }\textbf {\bibinfo {volume} {25}},\ \bibinfo {pages}
  {117106} (\bibinfo {year} {2016})}\BibitemShut {NoStop}%
\bibitem [{\citenamefont {Pershoguba}\ \emph {et~al.}(2018)\citenamefont
  {Pershoguba}, \citenamefont {Banerjee}, \citenamefont {Lashley},
  \citenamefont {Park}, \citenamefont {\AA{}gren}, \citenamefont {Aeppli},\
  and\ \citenamefont {Balatsky}}]{DiracMagnon}%
  \BibitemOpen
  \bibfield  {author} {\bibinfo {author} {\bibfnamefont {S.~S.}\ \bibnamefont
  {Pershoguba}}, \bibinfo {author} {\bibfnamefont {S.}~\bibnamefont
  {Banerjee}}, \bibinfo {author} {\bibfnamefont {J.~C.}\ \bibnamefont
  {Lashley}}, \bibinfo {author} {\bibfnamefont {J.}~\bibnamefont {Park}},
  \bibinfo {author} {\bibfnamefont {H.}~\bibnamefont {\AA{}gren}}, \bibinfo
  {author} {\bibfnamefont {G.}~\bibnamefont {Aeppli}}, \ and\ \bibinfo {author}
  {\bibfnamefont {A.~V.}\ \bibnamefont {Balatsky}},\ }\href {\doibase
  10.1103/PhysRevX.8.011010} {\bibfield  {journal} {\bibinfo  {journal} {Phys.
  Rev. X}\ }\textbf {\bibinfo {volume} {8}},\ \bibinfo {pages} {011010}
  (\bibinfo {year} {2018})}\BibitemShut {NoStop}%
\bibitem [{Dir(1994)}]{DiracMagnon2}%
  \BibitemOpen
  \href {\doibase https://doi.org/10.1016/0550-3213(94)90410-3} {\bibfield
  {journal} {\bibinfo  {journal} {Nuclear Physics B}\ }\textbf {\bibinfo
  {volume} {424}},\ \bibinfo {pages} {595 } (\bibinfo {year}
  {1994})}\BibitemShut {NoStop}%
\bibitem [{\citenamefont {Elias}\ \emph {et~al.}(2011)\citenamefont {Elias},
  \citenamefont {Gorbachev}, \citenamefont {Mayorov}, \citenamefont {Morozov},
  \citenamefont {Zhukov}, \citenamefont {Blake}, \citenamefont {Ponomarenko},
  \citenamefont {Grigorieva}, \citenamefont {Novoselov}, \citenamefont
  {Guinea},\ and\ \citenamefont {et~al.}}]{DiracMagnon3}%
  \BibitemOpen
  \bibfield  {author} {\bibinfo {author} {\bibfnamefont {D.~C.}\ \bibnamefont
  {Elias}}, \bibinfo {author} {\bibfnamefont {R.~V.}\ \bibnamefont
  {Gorbachev}}, \bibinfo {author} {\bibfnamefont {A.~S.}\ \bibnamefont
  {Mayorov}}, \bibinfo {author} {\bibfnamefont {S.~V.}\ \bibnamefont
  {Morozov}}, \bibinfo {author} {\bibfnamefont {A.~A.}\ \bibnamefont {Zhukov}},
  \bibinfo {author} {\bibfnamefont {P.}~\bibnamefont {Blake}}, \bibinfo
  {author} {\bibfnamefont {L.~A.}\ \bibnamefont {Ponomarenko}}, \bibinfo
  {author} {\bibfnamefont {I.~V.}\ \bibnamefont {Grigorieva}}, \bibinfo
  {author} {\bibfnamefont {K.~S.}\ \bibnamefont {Novoselov}}, \bibinfo {author}
  {\bibfnamefont {F.}~\bibnamefont {Guinea}}, \ and\ \bibinfo {author}
  {\bibnamefont {et~al.}},\ }\href {\doibase 10.1038/nphys2049} {\bibfield
  {journal} {\bibinfo  {journal} {Nature Physics}\ }\textbf {\bibinfo {volume}
  {7}},\ \bibinfo {pages} {701–704} (\bibinfo {year} {2011})}\BibitemShut
  {NoStop}%
\end{thebibliography}

\end{document}